\documentclass[twocolumn, prx, aps, superscriptaddress, longbibliography, amsmath, amssymb, floatfix]{revtex4-1}
\usepackage[colorlinks,linkcolor=blue,anchorcolor=blue,citecolor=blue,urlcolor=blue]{hyperref}
\usepackage{graphicx}
\usepackage{mathrsfs}
\usepackage{color}
\usepackage{mathtools}
\usepackage{bm}
\usepackage{bbm}
\usepackage{physics}
\renewcommand{\vec}[1]{\bm{#1}}
\begin{document}
\title{Relativistic Path-Integral Origin of the Dirac Equation, Quantum Collapse, Decoherence and Non-Hermitian Phenomena}
\author{Wei Wen}
\email{wenwei@hut.edu.cn}
\affiliation{College of Science, Hunan University of Technology, Zhuzhou 412007, Hunan Province, China}

\date{\today }
\begin{abstract}
Relativity and quantum mechanics are two cornerstones of modern physics, yet their unification within a single-particle path integral and a dynamical explanation of quantum measurement remain unresolved. Historically, these two problems have been treated as separate, but here we show they are intimately linked. We construct a self-consistent relativistic path integral that yields the Dirac and other standard wave equations under differetialable potentials. More importantly, we find that this propagator contains a latent, nonlocal correlation that is activated by realistic electromagnetic noise. This correlation unifies unitary evolution and wave-function collapse into a single dynamical mechanism: while differentiable potentials preserve unitary driving, nondifferentiable noise activates a bounded-martingale stochastic process that induces collapse. We show that the characteristics of quantum measurement are naturally derived from this stochastic dynamical process, thereby turning the axioms of quantum measurement from postulates into dynamical consequences. Furthermore, averaging this stochastic evolution over the noise record recovers the Gorini-Kossakowski-Sudarshan-Lindblad (GKSL) master equation, providing a first-principles derivation of decoherence free from the method of Born-Markov approximation. Extending this approach to composite systems establishes a stochastic foundation for effective non-Hermitian descriptions while preserving relativistic causality. Finally, because the noise spectrum governs the collapse process, engineering ``colored'' noise can actively accelerate or steer state reduction, suggesting new routes toward fast qubit reset and enhanced quantum control.
\end{abstract}
\maketitle

\section*{Introduction}
For decades, the foundations of quantum mechanics have been shadowed by a series of persistent, seemingly disparate puzzles. While Feynman's path integral connects the classical action to the Schr\"odinger equation \cite{FeynmanHibbs1965, Kleinert2009}, formulating a corresponding formula that bridges the classical relativistic action directly to the Dirac equation has remained an open mathematical problem \cite{Schulman1981, PeskinSchroeder1995}. Alongside this, another theoretical gap lies the measurement problem \cite{Bell1990_AgainstMeasurement, Zurek2003, Schlosshauer2005}. Historically, unitary evolution and wave-function collapse have been treated as fundamentally distinct dynamical rules. Yet, since both are ultimately driven by external potentials, what determines whether a given physical operation results in coherent evolution or irreversible collapse? Furthermore, if collapse is intrinsically stochastic, what microscopic mechanism generates this randomness without violating relativistic causality \cite{AharonovAlbert1981, Gisin1989,Bassi2013}? These core questions, regarding the universality and precision of standard decoherence theory (which relies on foundations such as the Born-Markov approximation \cite{Breuer2002}) and the causality paradoxes in effective non-Hermitian models \cite{Ashida2020}, have largely been treated within disconnected theoretical regions. These questions appear as scattered fragments of a puzzle, leaving us without a coherent picture and impeding our deeper understanding of quantum mechanics. This naturally leads us to ask if there is a minimal missing piece that allows us to assemble these fragments into a complete picture, thereby resolving these questions within a unified theoretical dynamics?

Existing approaches address specific aspects of these problems but lack a common mechanism. To circumvent the integration difficulties of the relativistic action, past works have modified the standard path-integral formulation. Early approaches discretize the Dirac propagator into a 1+1 dimensional zig-zag (checkerboard) sum \cite{FeynmanHibbs1965, feynman2010quantum, TJacobson_1984, ORD1993244, Earle1996}, while subsequent methods introduce auxiliary structures---such as Grassmann variables or proper-time constraints---to linearize the action and incorporate spin degrees of freedom \cite{BrinkDiVecchiaHowe1977NuclPhysB, BerezinMarinov1977AnnPhys, Schwinger1951, fradkin1991path, CORRADINI2021115498}. In parallel, the measurement problem has been approached through interpretational shifts \cite{Everett1957, bohm1952suggested} or by explicitly modifying the Schr\"odinger equation with non-linear stochastic terms, as seen in spontaneous collapse models \cite{bassi2003257, vonneumann1932mathematische}. At the macroscopic level, open-system theories describe the transition to classicality via environmental decoherence and master equations \cite{Zurek2003, Gorini1976, Lindblad1976, Breuer2002}, while quantum-trajectory and non-Hermitian formalisms describe conditioned evolution and dissipative dynamics \cite{Carmichael1993, WisemanMilburn2009, ElGanainy2018, Ashida2020}. However, treating these effective non-Hermitian models as unconditional laws often leads to physical inconsistencies, such as superluminal signaling \cite{PlenioKnight1998, dalibard1992wave}. Each work succeeds within its specific scope, but none simultaneously resolves the mystery of the missing relativistic path integral theory for single particles, provides a dynamical mechanism capable of identifying and distinguishing between unitary evolution and measurement collapse, or reproduces decoherence theory and non-Hermitian mechanisms without introducing additional physical postulates.

In this work, we find that the mathematical structure required to construct a covariant and analytically closed single-particle relativistic path integral is precisely this minimal missing piece. It not only assists us in establishing a unified path-integral theory but also allows the remaining questions to be resolved within the same theoretical dynamics. Achieving this closed form has historically been hindered by a big obstacle: relativistic actions generate non-quadratic phase factors, meaning that the velocity integrals that are Gaussian in the nonrelativistic limit cannot be solved in a straightforward analytic manner. We resolve this analytical closure problem---the explicit analytic propagator for relativistic particles---while remaining Feynman’s original thought. We provide a generalized path integral formula that is compatible with Feynman's path integral theory. With this formula, the previously non-Gaussian integrals can be calculated analytically through Bessel-function identities. The resulting propagator yields the Dirac equation for spin-$1/2$ particles \cite{Dirac1928a,dirac1981principles} and the square-root Hamiltonian form for scalar particles.

More important, this formula also provides a dynamical origin for quantum measurement and decoherence. We demonstrate that unitary evolution and wave-function collapse are two regimes of the same relativistic path-integral dynamics, determined by the differentiability of the external electromagnetic potential. In smooth potentials, a latent nonlocal contribution within the propagator is cancelled by the covariant transport, resulting in standard local, unitary evolution. However, in the presence of non-differentiable environmental electromagnetic noise, this cancellation is incomplete. The residual term in the propagator kernel drives a continuous stochastic evolution. In the eigenbasis of the unperturbed system Hamiltonian, the resulting probability dynamics follow a bounded-martingale stochastic differential equation \cite{williams1991probability,Doob1953,Oksendal2013}. The mathematical properties of this bounded martingale map directly to the measurement postulates: the absorbing boundaries define the preferred basis, the absorption probabilities yield Born's rule, and the mean first-passage time establishes a finite physical timescale for the collapse \cite{gardiner1985handbook,risken1996fokker}. 

Furthermore, averaging this stochastic evolution over the noise record recovers the GKSL master equation \cite{Gorini1976,Lindblad1976} without relying on the traditional Born-Markov weak-coupling derivation from an explicit system-bath model \cite{Breuer2002}. Beyond this, by extending this framework to multipartite entangled systems, we find that the conditional evolution of the global state naturally generates effective non-Hermitian generators that are widely used to describe conditional dissipation and monitored open systems. In our theory, these non-Hermitian structures are not introduced phenomenologically but from conditioning on environmental fluctuations. Equally important is that these non-Hermitian terms exist only at the level of conditioned realizations. After averaging over the noise record, the evolution reduces to an ordinary completely positive trace-preserving map, and no-signaling remains strictly preserved \cite{Dalibard1992,PlenioKnight1998,Ashida2020}. In this way, our theory places effective non-Hermitian dynamics and relativistic causal protection within a single dynamical equation. 

The paper is organized as follows. Section \ref{sec:dirac} outlines the construction of the relativistic path integral, showing how it recovers the Dirac, fractional square-root, Schr\"odinger, and Klein-Gordon equations. Section \ref{sec:measurement} derives the noise-activated stochastic propagator and the resulting bounded-martingale dynamics, calculating the average collapse time via first-passage theory. Section \ref{sec:decoherence} demonstrates how the GKSL equation arises as the ensemble average of these collapse dynamics. Section \ref{sec:nonhermitian} generalizes the framework to bipartite entangled systems, deriving effective non-Hermitian dynamics induced by local noise while maintaining complete positivity. Finally, Section \ref{sec:colored_noise} analyzes the dynamics under colored noise spectra. We detail the mechanisms by which colored noise accelerates or suppresses wavefunction collapse and how long-memory noise biases the Born rule, while also discussing the significant application potential of this work in the fields of rapid quantum state preparation and decoherence preservation technologies.

\section{Relativistic path-integral origin of relativistic evolution equations}
\label{sec:dirac}
It is often assumed that a single-particle relativistic path integral is either mathematically ill-posed or physically unnecessary. This view originates from two well-known difficulties. First, the relativistic action produces a non-quadratic phase factor, so the Gaussian techniques that make the nonrelativistic path integral analytically tractable no longer apply \cite{FeynmanHibbs1965,Schulman1981,Kleinert2009}. As a result, the relativistic kernel is frequently regarded as non-closable, Therefore, the relativistic kernel is often considered non-closable, meaning that no explicit analytical expression exists to bridge its classical relativistic action and the Dirac equation. Second, because particle number conservation is violated at high energies, it is widely argued that any single-particle description is inherently flawed, and therefore, therefore, a relativistic path integral theory for single particles must be superseded by the path integral formulation within the framework of Quantum Electrodynamics (QED).

We argue that these conventional concerns, though historically justified, do not invalidate the existence of single-particle approach. Regarding the first difficulty, the failure of Gaussian integration merely indicates the need for an alternative mathematical method. It is a mathematical difficulty rather than a physical prohibition. Regarding the second, while QFT is the proper framework for multi-particle creation and annihilation, this fact does not eliminate the necessity of the single-particle dynamical spacetime theory. Fundamentally, a path integral sums over all possible spacetime trajectories, including those that are spacelike. Even in low-energy regimes where particle creation is negligible, the relativistic effects of such contributions may still significantly affect the quantum propagator. For this reason, the relativistic effects of spacelike paths must be taken into account even when the particle itself is in a low-energy state.

In this work, we show that such a single-particle path integral does indeed exist. Under a specific covariant construction, the relativistic kernel becomes analytically closed in a dimensionally consistent manner, while still retaining the relativistic contribution of the underlying paths, and thereby establishing a direct link between the classical relativistic action and the Dirac equation.

\subsection{Construction of the Single-Particle Relativistic Path Integral}
\label{sec:diracA}
\begin{figure}[h!] 
    \centering 
    \includegraphics[width=0.49\textwidth]{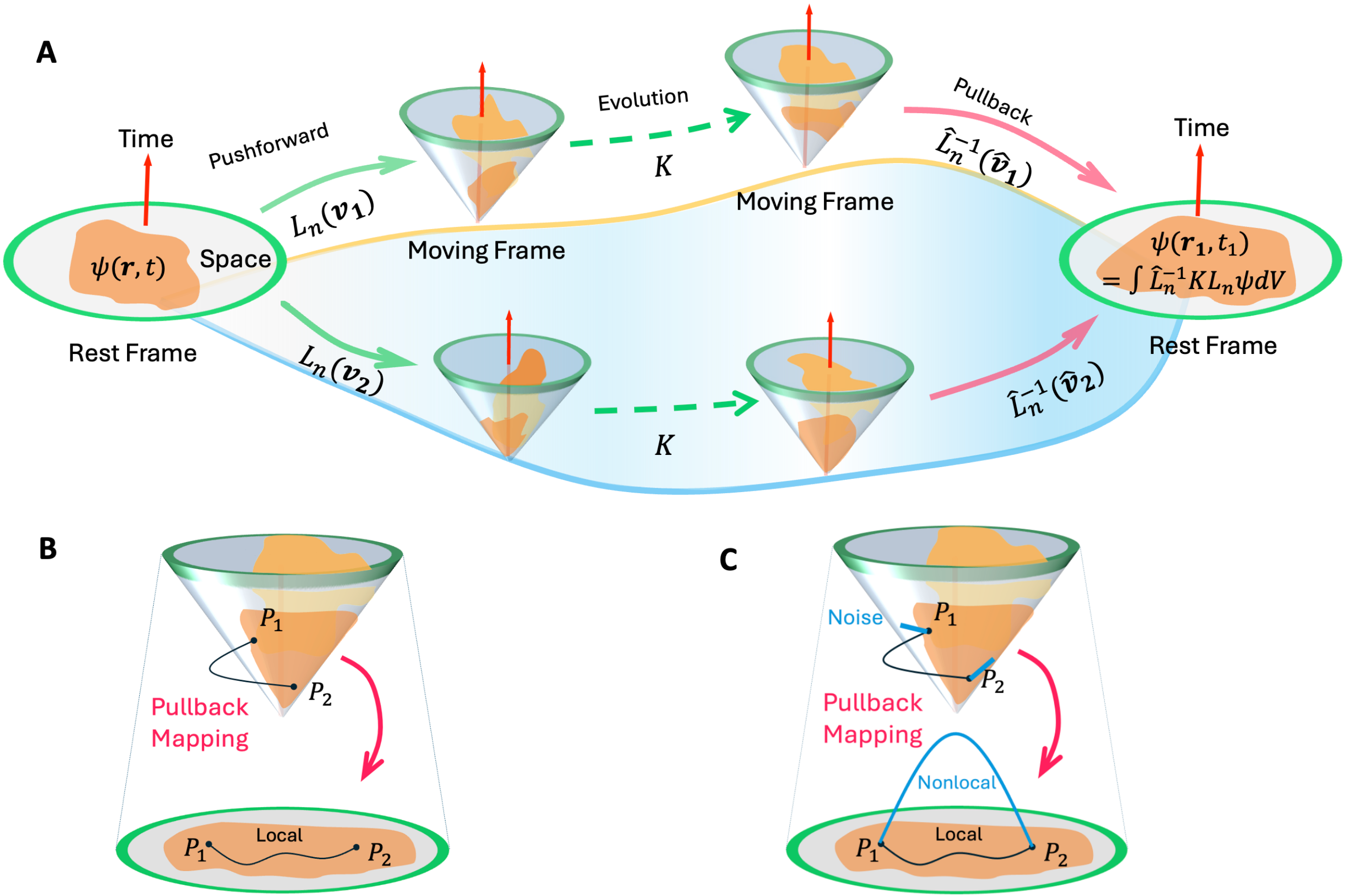} 
    \caption{\textbf{Relativistic pushforward -- pullback mechanism and nonlocality.} \textbf{A) Schematic of the covariant evolution}. The laboratory equal-time slice (rest frame) and the particle's moving slice (moving frame, i.e. the particle's proper frame) are different hyperplanes in spacetime, so viewing the lab state on the moving slice introduces (i) a change of spatial volume element, (ii) a boosted spin frame, and (iii) an amplitude renormalization that preserves probability flux. Accordingly, the state is pushed forward to the moving frame by $L_n=F_B F_V^{n}F_\Psi$, evolved by $K$, and then pulled back by $\hat{L}_n^{-1}$ to yield the observable result in the rest frame: $\Psi(\vec{r},t)= \int \hat{L}_n^{-1} K  L_n  \Psi(\vec{r}_0,t_0)\mathrm{d}V_0$. \textbf{B) differentiable environments.} Although individual moving-frame histories may include superluminal segments (black curve outside the cone joining $P_1$ and $P_2$), the pullback mapping exactly cancels this intrinsic nonlocality, and the image on the rest-frame space is a local connection between $P_1$ and $P_2$. \textbf{C) Non-differentiable (noisy) environments}. Random, non-differentiable noise (blue path) disrupts the smooth geometry of the mapping. Geometrically, the rough path prevents the pushforward and pullback transformations from perfectly cancelling. This resulting geometric mismatch is precisely the residual holonomy, a finite nonlocal correction that becomes the physical kernel for quantum collapse.}
    \label{fig:framework} 
\end{figure}

Previous approaches often modify the action, introduce auxiliary fields, or employ proper-time reparametrization to render the relativistic kernel mathematically tractable \cite{Schwinger1951,Schubert2001,Strassler1992,BrinkDiVecchiaHowe1976}. In contrast to these methods, our approach preserves the classical relativistic action. We find that the primary mathematical obstacle to constructing a relativistic path integral lies not merely in the non-quadratic phase factor, but rather in how the quantum state, the spatial volume element, and the probability amplitude must consistently transform across different inertial frames along each infinitesimal segment of a path. Consider a particle propagating along a given spacetime path. To a laboratory observer, the physical quantities defined on an equal-time hypersurface appear Lorentz-distorted when viewed from the instantaneous rest frame associated with that moving path segment. Since the relativistic path integral sums over all such segments, a consistent mathematical construction must geometrically correct for this kinematic mismatch. Before suming the phase contribution of a segment, the laboratory state and its associated integration measure must be projected, or pushed forward, to the particle’s proper frame. After the segment evolves, the result must be pulled back to the laboratory frame to yield an observable physical prediction (see Fig.~\ref{fig:framework}A).

This geometric transport modifies the propagator by inserting a pushforward operator $L_n$ and a corresponding pullback operator $\hat L_n^{-1}$.  The covariant evolution law takes the form
\begin{equation}
	\Psi(\vec r,t)=\int\hat L_n^{-1}K(\vec r,t;\vec r_0,t_0)L_n\Psi(\vec r_0,t_0)\mathrm d^n \vec r_0 ,
\label{eq:Relativistic_PI_final_ultimate}
\end{equation}
where the kernel,
\begin{equation*}
	K(\vec r,t;\vec r_0,t_0) =C_0\sum_{\wp_j}\exp(\frac{\mathrm{i}}{\hbar}\int_{\vec{r}_0,t_0}^{\vec{r},t}\mathcal{L}\mathrm{d}t),	
\end{equation*}
retains the standard Feynman summation over spacetime paths. The normalization constant $C_0$ and the transport operators systematically account for the relativistic kinematic corrections required to maintain Lorentz covariance. The compatibility with the traditional Feynman path integral becomes transparent in the nonrelativistic limit. When $c \to \infty$, relativistic distortions vanish, and the transport operators reduce to the identity transformation: $L_n \to 1$, $\hat L_n \to 1$. Simultaneously, the normalization reduces to the familiar Gaussian prefactor, $C_0 \to c_0^{n/2}=(\frac{m_0}{2\mathrm i \pi \hbar (t-t_0)})^{n/2}$, ensuring that Eq.~\eqref{eq:Relativistic_PI_final_ultimate} smoothly recovers the conventional non-relativistic Feynman propagator \cite{Feynman1948}. Thus, the present construction seamlessly extends the standard theory into the relativistic domain without replacing its fundamental principles.

The pushforward ($L_n$) and pullback ($\hat L_n$) operators are generated by a universal master function $f(x)$,
\begin{equation*}
	L_n, \hat{L}_n \xrightarrow{\text{generated by}} f_n(x)=F_B(x)F_V^{n}(x)F_\Psi(x),
\end{equation*}
which incorporates three distinct relativistic effects:

\textbf{1) Boost transformation ($F_B$).} For spin-$1/2$ particles, the local spin frame must align with the path. For a velocity $\vec v$, the corresponding Lorentz boost representation is
\begin{equation*}
	F_B(\vec{\alpha}\cdot\vec v)=\exp\left[\tfrac12 \operatorname{arctanh}\left(\frac{\vec\alpha\cdot\vec v}{c}\right)\right],
\end{equation*}
where $\vec\alpha$ and $\beta$ are standard Dirac matrices satisfying$\{\alpha_j,\alpha_k\}=2\delta_{jk}$ and $\{\alpha_j,\beta\}=0$. The physical role of $F_B$ becomes explicit when acting on the Dirac representation of a spacetime four-vector, $X = ct + \vec\alpha\cdot\vec r$. Under action by $F_B$, one obtains $X' = F_B(\vec{\alpha}\cdot\vec v)\, X\, F_B(-\vec{\alpha}\cdot\vec v)$, which yields precisely the Lorentz-transformed components: $ct'=\gamma_L(ct+\frac{\vec v\cdot\vec r}{c})$, $\vec r'\cdot\vec{e}_v=\gamma_L(\vec r\cdot\vec{e}_v+  vt)$, $\vec r'\times \vec{e}_v=\vec r\times \vec{e}_v$, with $\gamma_L=(1-\vec{v}^2/c^2)^{-1/2}$. Thus $F_B$ implements the standard Lorentz boost in the spior representation  \cite{BjorkenDrell1964,PeskinSchroeder1995}.

\textbf{2) Half-density factor ($F_V^{n}$).} The laboratory normalization of the wavefunction involves the spatial volume element $\mathrm d^n\vec r$, which undergoes Lorentz contraction along the direction of motion. Meanwhile, relativistic probability flux must remain covariantly conserved. To reconcile these, the wavefunction must transform as a half-density, introducing the dimension-dependent Jacobian factor
\begin{equation*}
	F_V^{n}(\vec v)=\left[\frac{2\gamma_L}{1+\gamma_L}\right]^{n/2}.
\end{equation*}

\textbf{3) Amplitude rescaling ($F_\Psi$).} Lorentz time dilation modifies the relationship between proper-time and laboratory-time evolution amplitudes. This kinematic effect contributes the final factor
\begin{equation*}
	F_\Psi(\vec v)=\sqrt{\frac{1+\gamma_L}{2}}.
\end{equation*}

The specific operators $L_n$ and $\hat L_n$ are constructed by evaluating $f_n$ on the velocity variable $\vec v$ or the corresponding momentum-derived operator $\hat{\vec v}$. We can verify that each of these constituent factors explicitly approaches unity as $c\to\infty$. Because $\gamma_L\to1$ and $\vec\alpha\cdot\vec v/c \to 0$ in the classical limit, it directly follows that
\begin{equation*}
	F_B \to \mathbf 1,\qquad F_V^{n} \to 1,\qquad F_\Psi \to 1.
\end{equation*}
This mathematically confirms our earlier assertion: $L_n,\hat L_n\to \mathbf 1$. By inserting the standard relativistic action into Eq.~\eqref{eq:Relativistic_PI_final_ultimate} with this covariant structure in place, the previously non-quadratic kernel can be evaluated analytically. For spinor particles this yields the Dirac equation, and for scalar particles the square-root Hamiltonian equation, as detailed in the following subsection.

\subsection{Spinor and scalar evolution equations}
\label{sec:diracB}
With the covariant transport structure established, the remaining task is to specify the relativistic Lagrangian and evaluate the corresponding kernel in the time-slice limit.

\textbf{1) Spin-$\tfrac{1}{2}$ particle.} For a charged spin-$\tfrac{1}{2}$ particle in an external electromagnetic field, the Lagrangian is taken in the Dirac-linear form
\begin{equation*}
	\mathcal L_R=\beta m_0 c^2\left(\vec\alpha\cdot\frac{\vec v}{c}-1\right)-U(\vec r,t),
\end{equation*}
where $U=qV-q\vec A\cdot\vec v$ is the electromagnetic potential. The action along a given path $\wp_j$ is therefore $S_R=\int_{\wp_j}\mathcal L_R\mathrm dt_1$. It is worth noting that the square of the kinetic energy part of $\mathcal{L}_R$, given by $[\beta m_0 c^2(\vec\alpha\cdot{\vec v}/{c}-1)]^2=[-m_0c^2\sqrt{1-\vec{v}^2/c^2}]^2$. This implies that $\mathcal{L}_R$ is the spinor formulation of our classical Lagrangian. This linear structure ensures that the relativistic phase factor couples directly to the Dirac matrices. Correspondingly, the time-slice normalization becomes matrix-valued, $C_0=(\beta c_0)^{n/2}$. The transport operators introduced in Sec.~\ref{sec:diracA} now take the explicit forms
\begin{equation*}
	L_n=f_n(\vec\alpha\cdot\vec v), \ \ \hat L_n=f_n(\beta\,\vec\alpha\cdot\hat{\vec v}), \ \ \hat{\vec v}=\frac{\hat{\vec p}-q\vec A}{m_0},
\end{equation*}
where the master function $f_n$ was defined previously as $f_n=F_B F_V^{n} F_\Psi$. Substituting $\mathcal L_R$, $L_n$, and $\hat L_n$ into the covariant propagation formula \eqref{eq:Relativistic_PI_final_ultimate} and performing the time-slice limit, the non-Gaussian velocity integrals close analytically \cite{quantum7040059}. The resulting evolution equation is
\begin{equation}
	\mathrm i\hbar\frac{\partial\Psi}{\partial t}=\Big(\beta m_0 c^2+c\vec\alpha\cdot(\hat{\vec p}-q\vec A)+qV\Big)\Psi,
\label{eq:Dirac_final}
\end{equation}
which is precisely the Dirac equation  \cite{Dirac1928a,dirac1981principles}.

\textbf{2) Spin-0 particle.} For a scalar particle, the relativistic segment Lagrangian is taken in the standard proper-time form
\begin{equation*}
	\mathcal L_r=-\gamma_L^{-1} m_0 c^2-U(\vec r,t),
\end{equation*}
leading to the action $S_r=\int_{\wp_j}\mathcal L_r\,\mathrm dt_1$, accompanied by the scalar normalization $C_0=c_0^{n/2}$. Due to the absence of an internal spin frame, the transport operators reduce to the scalar projection (denoted by $\mathrm{Sc}[\cdot]$) of the associated spinor functions:
\begin{equation*}
	L_n(\vec v)=\mathrm{Sc}\left[f_n(\vec\alpha\cdot\vec v)\right], \qquad \hat L_n=L_n(\mathrm i\hat{\vec v}).
\end{equation*}
Carrying out the same time-slice evaluation yields the first-order evolution
\begin{equation}
	\mathrm i\hbar\frac{\partial\Psi}{\partial t}=\left(\sqrt{m_0^2c^4+c^2(\hat{\vec p}-q\vec A)^2}+qV\right)\Psi.
\label{eq:Scalar_final}
\end{equation}
which constitutes the fractional square-root Hamiltonian equation. Spinor and scalar wave dynamics therefore arise from the identical covariant path-integral foundation, differing mathematically only in the representation space of the transport operators.

\paragraph{Nonrelativistic limit.} In the low-energy regime characterized by $\langle c^2(\hat{\vec p}-q\vec A)^2\rangle\ll m_0^2c^4$, the square-root Hamiltonian then reduces to 
\begin{equation}
	\mathrm i\hbar\partial_t\Psi=\left(\frac{(-\mathrm i\hbar\nabla-q\vec A)^2}{2m_0}+qV+m_0c^2\right)\Psi.
\end{equation}
The rest-mass term contributes only a global dynamical phase $\exp(-\mathrm{i}m_0 c^2 t/\hbar)$. Factoring out this phase leaves the energy level structure and superposition unaffected, strictly recovering the Schr\"odinger equation.

\paragraph{Evolution equation for negative-energy states.} Furthermore, this geometric formalism consistently accommodates particle states with negative energy. Choosing the sign-reversed Lagrangian $\mathcal L'_r = \gamma_L^{-1}m_0c^2 - U$ and the corresponding normalization $C'_0 = (-c_0)^{n/2}$ yields the negative branch of the square-root Hamiltonian for the wavefunction $\Psi'$. Forming the symmetric and antisymmetric combinations $\Phi_\pm=(\Psi\pm\Psi')/\sqrt2$ mathematically eliminates the square-root operator and produces the second-order Klein-Gordon equation,
\begin{equation}
	(\mathrm i\hbar\partial_t-qV)^2\Phi_\pm=\left(m_0^2c^4+c^2(-\mathrm i\hbar\nabla-q\vec A)^2\right)\Phi_\pm,
\end{equation}
demonstrating that Dirac, square-root, Sch\"ordinger and Klein-Gordon dynamics share a common covariant path-integral origin.

\subsection{Uniqueness of the Single-Particle Path Integral Form}
\label{sec:diracC}
From a mathematical perspective, the bottleneck for the analytic closure of the relativistic path integral is the appearance of non-Gaussian core integrals. In contrast to the nonrelativistic kernel, where the quadratic phase closes by elementary Gaussian integration, the phase of the relativistic kernel during a single time-slice evolution contains the proper-time factor $\sqrt{1-v^2/c^2}$. Consequently, utilizing the path integral to compute wave-function evolution inevitably involves the following complex integral:
 \begin{equation}
	\int_{-\infty}^{\infty} f_n(\vec{v}) e^{\pm \mathrm{i}\chi\sqrt{1-\frac{v^2}{c^2}}} \Psi(\vec{r}_0 + \vec{v}\epsilon, t) \mathrm{d}^n\vec{v},
\label{eq:initial_form}
\end{equation}
rendering the analytic closure of the relativistic path integral analytically intractable. Therefore, the problem of establishing the single-particle relativistic path integral formulation translates to: what functional form of $f_n(\vec{v})$ permits the analytic closure of the above integral and ultimately yields the unitary operator $\exp(-\mathrm{i}\hat{H}\epsilon/\hbar)$?

\paragraph{Spacelike paths and analytic continuation.}
For Eq.~\eqref{eq:initial_form} to serve as a mathematically valid form for wave-function evolution, the integral must remain convergent for every square-integrable wave function. In other words, the kernel $K=C_0\sum e^{\mathrm{i}S/\hbar}$ must not diverge. We note that as the trajectories extend into the spacelike region $|\vec v|>c$, the factor $\sqrt{1-v^2/c^2}$ becomes imaginary, converting the oscillatory phase factor $\exp(\mathrm{i}S/\hbar)$ into a real exponential. In this regime, instead of producing oscillatory cancellation, the integrand diverges exponentially along the real integration contour. Consequently, the action of the kernel on normalizable wave functions becomes ill-defined.

To preserve mathematical well-posedness, the integration contour must therefore be deformed by analytic continuation, ensuring that the kernel remains an integrable distribution and that the full propagator acts boundedly on physical states. The unique method to fulfill this mathematical requirement is that we must prescribe:
\begin{equation}
\mathcal{L}_{R|r}=\mathrm{i}m_0c^2\sqrt{\frac{v^2}{c^2}-1}-U,\qquad |\vec v|>c.
\end{equation}
Only through this prescription can the kernel serve as a legitimate propagator within the thought of Feynman's path integral.

\paragraph{Bessel identity and closure gate.}
With well-posedness ensured, the remaining problem is closure: whether the nonquadratic relativistic slice integral can be evaluated exactly and reorganized into $\exp(-\mathrm{i}\hat H\epsilon/\hbar)$. We find that the key to solving this problem lies in a crucial Bessel integral identity (Gradshteyn \& Ryzhik, \emph{Table of Integrals, Series, and Products}, 7th ed., 2007, p.~710, Eq.~6.646.1  \cite{Gradshteyn2007}):
\begin{equation}
\label{eq:GR_Bessel}
	\begin{aligned}
		&\int_1^\infty\left(\frac{x-1}{x+1}\right)^{\mu/2}e^{-\alpha x}j_{\mu}\bigl(\beta\sqrt{x^2-1}\bigr)\mathrm{d}x \\
		=&\frac{1}{\sqrt{\alpha^2+\beta^2}}\left(\frac{\beta}{\alpha+\sqrt{\alpha^2+\beta^2}}\right)^{\mu}e^{-\sqrt{\alpha^2+\beta^2}} .
	\end{aligned}
\end{equation}
We refer to this as the ``Closure Gate''. After the standard hyperbolic change of variables that converts the velocity integral into an integral over $x\in[1,\infty)$, the relativistic phase generates the Bessel argument $\sqrt{x^2-1}$, while the covariant slice measure together with the transport prefactor produces precisely the algebraic weight $\big(\frac{x-1}{x+1}\big)^{\mu/2}$. The exponential term on the right-hand side of \eqref{eq:GR_Bessel} then reproduces the evolution operator $\exp(-\mathrm{i}\hat H\epsilon/\hbar)$, thereby providing an explicit analytic bridge from the relativistic action to the Hamiltonian generator (see Appendix \ref{app:evolution_from_timeslice}).

\paragraph{Uniqueness of the transport operators.}
Equation~\eqref{eq:GR_Bessel} reveals a stringent structural constraint: the closure gate opens only if the integrand carries the weight $\big(\frac{x-1}{x+1}\big)^{\mu/2}$ with the correct order $\mu$, which is determined by the spatial dimension and by whether one extracts diagonal or spin-coupling components. Within the path integral the parameter $\mu$ is therefore not adjustable; it is locked by the dimension of the spatial measure and by the algebraic structure of the transported state. Requiring that the same relativistic one-slice kernel close consistently across $n=1,2,3$ spatial dimensions removes any residual freedom in defining the transport operators $(L_n,\hat L_n)$.

In particular, the prefactor must simultaneously satisfy:
(i) Lorentz covariance of transported states,
(ii) dimension-uniform analytic closure via Eq.~\eqref{eq:GR_Bessel}, and
(iii) the nonrelativistic limit $c\to\infty$ with $L_n,\hat L_n\to 1$. 

We find that only under the following construction 
\begin{equation*}
	f_n = F_B\,F_V^{\,n}\,F_\Psi,\ \ L_n(\vec{v})=\mathrm{Sc}(f_n(\vec{\alpha}\cdot\vec v)),\ \ \hat L_n=L_n(\mathrm{i}\hat{\vec v}).
\end{equation*}
does the Eq.~\eqref{eq:initial_form} satisfy the above requirements and make
\begin{align}
	&\int_{-\infty}^{\infty} f_n(\vec{v}) e^{\pm \mathrm{i}\chi\sqrt{1-\frac{v^2}{c^2}}} \Psi(\vec{r}0 + \vec{v}\epsilon, t) \mathrm{d}^n\vec{v} \nonumber \\
\sim &\int_1^\infty\left(\frac{x-1}{x+1}\right)^{\frac{n-2}{4}}e^{-\alpha x}j_{\mu}\bigl(\beta\sqrt{x^2-1}\bigr)\mathrm{d}x \sim e^{-\mathrm{i}\hat{H}_r\epsilon/\hbar}. \nonumber
\end{align}
Any alteration modifies the induced power of $\frac{x-1}{x+1}$ and breaks the analytic match to Eq.~\eqref{eq:GR_Bessel}, thereby obstructing analytic closure. Consequently, the form of the transport operators in this paper possesses uniqueness; what we can do is provide a physical interpretation for this mathematical form.

\paragraph{Uniqueness of the spinor action form.} Once the scalar kernel is analytically determined by the Bessel formulae, the spinorial extension is determined by Lorentz representation theory. A spin-$1/2$ state transforms under the spinor representation of the Lorentz group, which requires a mathematical match between the time-slice phase and the spinor boost $F_B(\vec v)$. This necessity explains the emergence of the Dirac-linear segment Lagrangian $\mathcal{L}_R=\beta m_0c^2(\vec{\alpha}\cdot\vec{v}/c-1)-U(\vec{r},t)$. It is the only algebraic form whose phase factor admits the symmetric factorization
\begin{equation*}
	F_B(\vec{\alpha}\cdot\vec v)e^{\tfrac{\mathrm{i}}{\hbar}\mathcal{L}_R\epsilon}F_B(-\vec{\alpha}\cdot\vec v)= e^{\tfrac{\mathrm{i}}{\hbar}\beta \mathcal{L}_r\epsilon}.
\end{equation*}
This factorization allows the spinorial kernel to reduce component-wise to the previously established scalar Bessel closure. 

In summary, the present path-integral formulation originates from a series of mathematical and physical requirements: the convergence of the propagator along spacelike paths, the compatibility of path-integral physical quantities across hypersurfaces, and the dimension-uniform analytic closure mandated by the Bessel identity Eq.~\eqref{eq:GR_Bessel}. Together, these requirements render the transport prefactor $f_n=F_BF_V^{\,n}F_\Psi$ and its induced propagator form unique.

\section{Relativistic Path-integral Origin of Quantum Measurement}
\label{sec:measurement}
\subsection{Noise-activated collapse dynamics}
One of our central findings is that unitary evolution and quantum collapse are not governed by distinct physical laws, but are two operational regimes of the same relativistic equation \eqref{eq:Relativistic_PI_final_ultimate}. The distinction is entirely determined by the differentiability of the environmental potential.

To see how this distinction arises, we evaluate the evolution over an infinitesimal time interval $\epsilon \to 0$. In this limit, the path-integral summation is dominated by highly spacelike trajectories ($|\dot{\vec{r}}| \gg c$). Consequently, the classical actions reduce to:
\begin{equation*}
	S_{R|r} \xrightarrow{\epsilon\to 0} \mathrm{i} m_0 c |\Delta\vec{r}| + q\vec{A}\cdot\Delta\vec{r}.
\end{equation*}
This implies that as $\epsilon\rightarrow 0$, the phase factor $e^{\mathrm{i}S_{R|r}/\hbar}$ no longer approaches the constant $1$. More important, the phase term $-q\vec{A}\cdot\Delta\vec{r}$ implies that a sudden fluctuation in the vector potential $\vec{A}$ is instantaneously correlated across spatial distances $\Delta\vec{r}$. This exposes a key feature: the relativistic propagator defined by Eq.~\eqref{eq:Relativistic_PI_final_ultimate} possesses an intrinsic, latent nonlocality. This underlying nonlocal correlation is the fundamental origin of the bifurcation between unitary evolution and wave-function collapse.

Because the spinor path integral contains a richer structure than the scalar one, we use it as an illustrative example below. We define $K_L(\vec{r}_0+\Delta\vec{r},t+\epsilon;\vec{r}_0,t)=\hat{L}_n^{-1}K L_n$, and to distinguish it from the bare kernel $K$, we refer to it as the covariant propagator. It is important to note here that because $\hat{L}_n$ is a function of the momentum operator $\hat{\vec{P}}=-\mathrm{i}\hbar\nabla_{\vec{r}}-q\vec{A}(\vec{r})$, it manifests in the coordinate representation as a differential operator acting on the endpoint coordinate $\vec{r}=\vec{r}_0+\Delta\vec{r}$. Within the coordinate representation of the propagator, $\hat{L}_n$ operates solely by differentiating the first argument of the kernel function $K(\vec{r}_0+\Delta\vec{r},t+\epsilon;\vec{r}_0,t)$. It does not act on the initial wave function $\Psi(\vec{r}_0,t_0)$ because $\vec{r}$ is independent of $\vec{r}_0$. Consequently, the evaluated expression $\hat{L}_n^{-1}K L_n$ remains a function rather than an operator. Using the explicit spinor forms of $\hat{L}_n$ and $L_n$, the instantaneous covariant propagator evaluates to:
\begin{equation*}
	\begin{aligned}
		&K_L(\vec{r}_0+\Delta\vec{r},t^+;\vec{r}_0,t):=\hat{L}_n^{-1}K(\vec{r}_0+\Delta\vec{r},t^+;\vec{r}_0,t) L_n  \\
		&=\hat{L}_n^{-1} \left[ \left(\frac{-m_0 c\beta}{2\pi\hbar|\Delta\vec{r}|}\right)^{n/2} e^{\mathrm{i}\beta\vec{\alpha}m_0 c|\Delta\vec{r}|/\hbar} e^{\mathrm{i}q\vec{A}\cdot\Delta\vec{r}/\hbar} \right].
	\end{aligned}
\end{equation*}
We observe that the behavior of the covariant kernel depends on the environment:

\textbf{1) Differentiable environments (Unitary limit):} When the potentials $\vec{A}$ and $V$ are smooth, continuous functions, the covariant transport cancels the spacelike nonlocality. The covariant propagator acts as a local Dirac delta distribution: $K_L(\vec{r}_0+\Delta\vec{r},t^+;\vec{r}_0,t) = \delta(\Delta\vec{r})$. The evolution remains local and deterministic, governed solely by the unitary Dirac equation (Fig.~\ref{fig:framework}B).
	
\textbf{2) Non-differentiable environments (Collapse limit):} When $\vec{A}$ and $V$ are non-differentiable (as is ubiquitous with quantum vacuum fluctuations or thermal noise), the equality $K_L = \delta(\Delta\vec{r})$ fails. The geometric roughness of the potential introduces a singularity in the pullback mapping $\hat{L}_n^{-1}$, preventing it from concealing the underlying spacelike trajectories. Geometrically, as illustrated in Fig.~\ref{fig:framework}C, this breakdown is interpreted as a defect in the effective connection governing parallel transport: smooth transport is ill-defined on a non-differentiable gauge potential. The non-differentiable noise physically disrupts this smooth mapping, leaving a residual holonomy that manifests as a finite nonlocal correction, which ultimately drives quantum collapse.

In reality, macroscopic detectors and the surrounding vacuum inevitably introduce non-smooth, randomly fluctuating electromagnetic potentials $V_{\text{noise}}(\vec{r},t)$ and $\vec{A}_{\text{noise}}(\vec{r},t)$. These enter the action directly, decomposing it into a noise-free part $S_0=\int_{\wp}(T-qV_0+q\vec{A}_0\cdot\vec{v})\mathrm{d}t$ and a stochastic correction $S_{\text{noise}} =\int_{\wp} (-q V_{\text{noise}} + q \vec{A}_{\text{noise}} \cdot \vec{v})\mathrm{d}t$. The bare kernel $K$ acquires a stochastic correction via $\exp[\mathrm{i} (S_0+S_{\text{noise}})/\hbar]$. These non-differentiable increments propagate into the operator $\hat{L}_n$, and thus into $K_L$. Consequently, the covariant propagator becomes (see Appendix \ref{app:covariant_propagator_perturb}):
\begin{align}
	& K_L(\vec{r}_0+\Delta\vec{r}, t^{+}; \vec{r}_0, t) \nonumber\\
	& =\delta(\Delta \vec{r})+\biggr(\frac{q\mathrm{d}\vec{A}_I\cdot(\hat{\vec{p}}_0-q\vec{A}_0)}{4m_0^2c^2\Lambda_n(\hat{H}_0)}+\frac{\mathrm{d}\vec{B}_I\cdot \vec{\mu}_s}{4m_0c^2\Lambda_n(\hat{H}_0)}\biggl)\delta(\Delta \vec{r}) \nonumber\\
	& =\delta(\Delta\vec{r}) + \left(\mathrm{d}\vec{A}_I \cdot \hat{\vec{N}} + \mathrm{d}\vec{B}_I \cdot \hat{\vec{M}}\right)\delta(\Delta\vec{r}). 
	\label{eq:Nondiff_pathintegra}
	\end{align}
Here, $\mathrm{d}\vec{A}_I=\vec{A}_{\text{noise}}(t^+)-\vec{A}_{\text{noise}}(t)$ and $\mathrm{d}\vec{B}_I=\vec{B}_{\text{noise}}(t^+)-\vec{B}_{\text{noise}}(t)$ are infinitesimal noise increments. Crucially, these non-differentiable increments do not vanish in the limit of zero time step in the same manner as smooth differentials. This non-vanishing property is the source of the finite nonlocal correction, which manifests physically as quantum collapse. In Eq.~\eqref{eq:Nondiff_pathintegra}, the factor
\begin{equation*}
	\Lambda_n(\hat{H}_0)=\frac{2m_0c^2\hat{H}_0+2m_0^2c^4}{n\hat{H}_0^2+m_0c^2\hat{H}_0},
\end{equation*}
depends on the noise-free Hamiltonian 
\begin{equation*}
	\hat H_0=\sqrt{m_0^2c^4+(\vec{\alpha}\cdot(\hat{\vec{p}}-q\vec{A}_0))^2},
\end{equation*}
meaning that $\hat{\vec N}$ and $\hat{\vec M}$ are inherently operator-valued functions. Acting on the Dirac delta, they generate a derivative-dressed distribution acting as a nonlocal phase kernel. The evolution $\Psi(\vec{r}, t^{+}) = \int K_L \Psi(\vec{r}_0,t) \mathrm{d}^n\vec{r}_0$ is equivalent to applying an infinitesimal stochastic operator: $\Psi(\vec{r},t^{+}) = (1 + \mathrm{d}\vec{A}_I \cdot \hat{\vec{N}} + \mathrm{d}\vec{B}_I \cdot \hat{\vec{M}}) \Psi(\vec{r},t)$. Since $\hat{\vec N}$ and $\hat{\vec M}$ commute with $\hat H_0$, the dynamics can be described in the eigenbasis $\{\varphi_j(\vec r)\}$ of $\hat{H}_0$. For $\Psi(\vec r,t)=\sum_j a_j(t) \varphi_j(\vec r)$, we obtain:
\begin{equation}
	a_j(t+)=  a_j(t)\left(1+\mathrm{d}\vec{A}_I \cdot \vec{N}_j+\mathrm{d}\vec{B}_I\cdot\vec{M}_j \right).
\label{eq:collapse_equation_deriv_revised_latex}
\end{equation}
where $\vec N_j$ and $\vec M_j$ are the eigenvalue vectors induced by $\hat{\vec N}$ and $\hat{\vec M}$ on $|\varphi_j\rangle$. To analyze probability dynamics, we consider the normalized squared amplitudes $p_j(t) \equiv |a_j(t)|^2 / \sum_k |a_k(t)|^2$. 
From Eq.~\eqref{eq:collapse_equation_deriv_revised_latex} one obtains the following It\^o diffusion process for $\{p_j\}$:
\begin{equation}
	\mathrm{d}p_j=2p_j\sum_{k\neq j}p_k\Big(\vec N_{j,k}\cdot\mathrm d\vec A_I+\vec M_{j,k}\cdot\mathrm d\vec B_I\Big),
\label{eq:collapse_equation_ito_form}
\end{equation}
with $\vec N_{j,k}=\vec N_j-\vec N_k$ and $\vec M_{j,k}=\vec M_j-\vec M_k$. To connect these increments to standard Wiener processes, we model the electromagnetic noise as white-noise-limited fluctuations with delta correlations. In the Weyl gauge ($V_{\rm noise}=0$), one has $\vec E_{\rm noise}=-\partial_t\vec A_{\rm noise}$. If the electric field is idealized as a Gaussian white-noise process satisfying
\begin{equation*}
	\langle E_{\text{noise},\mu}(t)E_{\text{noise},\nu}(t')\rangle=\sigma_E^2\delta_{\mu\nu}\delta(t-t'),
\end{equation*}
then $\vec A_{\rm noise}(t)$ is an integrated white-noise process, and its increment over one time slice may be represented as
\begin{equation*}
	\mathrm d\vec A_I=\sigma_E\,\mathrm d\vec W^{E},
\label{eq:dA_wiener_short}
\end{equation*}
with $\mathbb E[\mathrm{d}A_{I,\mu}]=0$ and $\mathrm{d}A_{I,\mu}\mathrm{d}A_{I,\nu}=\sigma_E^2\delta_{\mu\nu}\mathrm{d}t$. Here the minus sign from $\vec E_{\rm noise}=-\partial_t\vec A_{\rm noise}$ is absorbed into the definition of $\mathrm d\vec W^{E}$---the standard Wiener increment\cite{williams1991probability,grimmett2020probability}.

The magnetic term requires a physical coarse-graining because the time derivative of ideal white noise is not defined. We therefore introduce an effective inverse timescale using the zero-crossing rate, which provides a natural frequency scale for a fluctuating signal  \cite{rice1944mathematical,CramerLeadbetter1967}. In our setting the corresponding scale $Z_{cr}={ \pi^2 h c^2 \epsilon_0^2 \sqrt{k_B Tm_0c^2}}/{q^4}$ is derived in Appendix \ref{app:zcr}. This leads to the effective relation $\mathrm d\vec B_I\sim Z_{\rm cr}\sigma_B\mathrm d \vec W^B$, and Eq.~\eqref{eq:collapse_equation_ito_form} becomes
\begin{equation*}
	\mathrm dp_j=2p_j\sum_{k}p_k\Big({\sigma}_E \vec N_{j,k}\cdot\mathrm d \vec W^E+Z_{\rm cr}{\sigma}_B\vec M_{j,k}\cdot\mathrm d \vec W^B\Big).
\end{equation*}
For notational convenience, it is useful to combine the electric and magnetic noise channels into a single index $\mu$. We therefore define
\begin{equation}
	(\mathrm dW_{\mu},\hat J_{\mu})=
	\begin{cases}
		(\mathrm dW_i^{E},\,\sigma_E \hat N_i), & \mu=(E,i),\\[3pt]
		(\mathrm dW_i^{B},\,Z_{\rm cr}\sigma_B \hat M_i), & \mu=(B,i).
	\end{cases}
\label{eq:new_definition}
\end{equation}
If $\{|\varphi_j\rangle\}$ is the common eigenbasis of the noise-free Hamiltonian and the commuting noise operators, we write
\begin{equation*}
	\hat J_{\mu}|\varphi_j\rangle = J_{\mu,j}|\varphi_j\rangle,
	\qquad
	J_{\mu}^{(j,k)}:=J_{\mu,j}-J_{\mu,k}.
\end{equation*}
Then Eq.~\eqref{eq:collapse_equation_ito_form} can be rewritten compactly as
\begin{equation}
	\mathrm dp_j=2p_j\sum_{k}\sum_{\mu} p_k\, J_{\mu}^{(j,k)}\,\mathrm dW_{\mu}.
\label{eq:Collapse_SDE_final_concise}
\end{equation}

\begin{figure*}[htbp]
\centering
\includegraphics[width=0.98\textwidth]{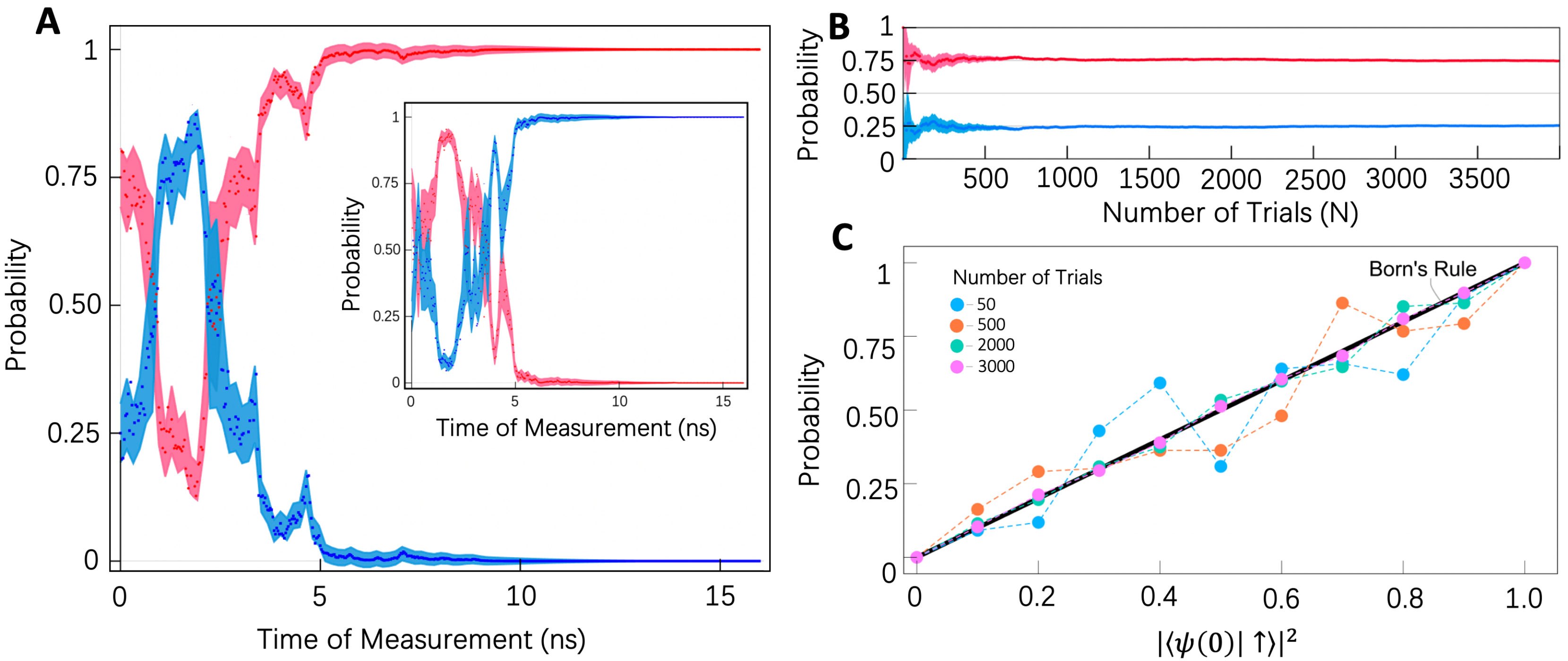}
\caption{\textbf{Noise-induced  wavefunction collapse and Born's rule.}
The collapse equation derived from our single-particle relativistic path integral, Eq.~\eqref{eq:Collapse_SDE_final_concise}, is a bounded-martingale SDE that naturally yields stochastic collapse and Born's rule. In this figure, we simulate a spin-$1/2$ two-level system at $T=300~\mathrm{K}$ with magnetic-field noise $\sigma_B=1.25~\mu\mathrm{T}/\sqrt{\mathrm{Hz}}$. \textbf{A) Time traces for the initial state $|\psi(0)\rangle=\frac{\sqrt{3}}{2}|\uparrow\rangle+\frac{1}{2}|\downarrow\rangle$}. For each tick of $1~\mathrm{ns}$ on the horizontal axis we generate $10^{4}$ micro-samples. The shaded bands depict the oscillation envelope spanned by these micro-samples and the dots in bands are block averages over 500 consecutive micro-samples; the red (blue) series corresponds to $p_\uparrow(t)$ ($p_\downarrow(t)=1-p_\uparrow(t)$). Because Eq.~(\ref{eq:Collapse_SDE_final_concise}) is a bounded martingale, trajectories are absorbed at $p_\uparrow=1$ or $0$ (collapse to $|\uparrow\rangle$ or $|\downarrow\rangle$); the inset shows the opposite outcome under a different noise realization. \textbf{B) Monte-Carlo convergence of collapse's results with the number of trials $N$}. The running estimates $P_\uparrow(N)$ (red) and $P_\downarrow(N)$ (blue) approach the initial weights $p_\uparrow(0)$ and $p_\downarrow(0)$, respectively. \textbf{C) Monte-Carlo validation of Born's rule.} For many choices of $|\psi(0)\rangle$, the fraction of collapses to $|\uparrow\rangle$ (points; colors denote independent batches) plotted against $|\langle\uparrow|\psi(0)\rangle|^{2}$ follows the Born-rule prediction (black line), $P_\uparrow(\infty)=p_\uparrow(0)=|\langle\uparrow|\psi(0)\rangle|^{2}$ and $P_\downarrow=1-P_\uparrow$.}
\label{collapsfig}
\end{figure*}
 
\subsection{From dynamics to the axioms of measurement}
Equation~\eqref{eq:Collapse_SDE_final_concise} describes a continuous stochastic evolution of the occupation probabilities $p_j(t)$ under environmental noise. Since the equation contains no deterministic drift term, each $p_j(t)$ is a bounded martingale\cite{williams1991probability,grimmett2020probability}. This simple mathematical structure can naturally account for the main features of quantum measurement.

First, the noise amplitude vanishes when one of the probabilities reaches $1$ and all others reach $0$. These eigenbasis $\{|\varphi_j\rangle\}$ of $\hat{H}_0$ are therefore absorbing boundaries: once the system reaches such a state, the stochastic evolution stops and the outcome becomes stable. Second, because the process is a bounded martingale, the probability of ending at a given absorbing boundary, $|\varphi_j\rangle$, is fixed by the initial value of the corresponding component. This directly yields Born's rule. Third, the approach to an absorbing boundary is a first-passage process, so the collapse time is not an additional postulate but a dynamical quantity that can be calculated from the associated backward equation.

Consequently, the core postulates of quantum measurement no longer require invoking external observers or hypotheses; instead, they follow directly as mathematical theorems of our work.

\textbf{1) Preferred basis and irreversibility.}  
The preferred basis is selected by the structure of the noise coupling. As shown in Eq.~\eqref{eq:Nondiff_pathintegra}, the noise operators $\hat{\vec N}$ and $\hat{\vec M}$ are constructed from the noise-free Hamiltonian $\hat H_0$, and therefore commute with it. As a result, the stochastic evolution is diagonal in the basis that simultaneously diagonalizes $\hat H_0$, $\hat{\vec N}$, and $\hat{\vec M}$. In the nondegenerate case this is simply the eigenbasis $\{|\varphi_j\rangle\}$ of $\hat H_0$; more generally, one should speak of the common eigenspaces selected by the Hamiltonian and the noise coupling. In this basis, the noise drives only the occupation probabilities $p_j(t)$ and does not generate new coherent superpositions between different basis states.

The stability of the final outcomes follows directly from the structure of the bounded-martingale SDE. In Eq.~\eqref{eq:Collapse_SDE_final_concise}, the diffusion term for the $j$-th component is proportional to $p_j\sum_{k\neq j}p_k(\cdots)$. It therefore vanishes when one probability reaches $1$ and all others vanish. These pure-state vertices are the absorbing states of the stochastic dynamics. Once a trajectory reaches such a vertex, the noise term disappears and the system no longer diffuses away from that state. In this sense, the preferred basis is not imposed externally, but is selected dynamically by the combined structure of the system Hamiltonian and its noise coupling.

\textbf{2) Born's rule and intrinsic randomness.}  
Within this theory, Born's rule follows directly from the bounded-martingale structure of the probability dynamics, while the randomness of a measurement outcome is traced to the particular realization of the environmental Wiener process.

Because Eq.~\eqref{eq:Collapse_SDE_final_concise} contains no deterministic $\mathrm dt$ drift term, each $p_j(t)$ is a bounded martingale. Let $\tau$ denote the absorption time at which the trajectory reaches one of the absorbing vertices. By the Optional Stopping Theorem, one has $\mathbb E[p_k(\tau)] = p_k(0)$ \cite{williams1991probability,Oksendal2013}. At the absorption time, the system has reached a definite outcome, so $p_k(\tau)$ can only take the values $0$ or $1$. The event $p_k(\tau)=1$ means that the state has collapsed onto the basis state $|\varphi_k\rangle$. Therefore, $\mathbb E[p_k(\tau)] = \mathbb P(\text{outcome }|\varphi_k\rangle)$, and hence
\begin{equation}
	\mathbb P(\text{outcome }|\varphi_k\rangle)=p_k(0)=|\langle\varphi_k|\Psi(0)\rangle|^2.
\label{eq:born_derived}
\end{equation}
This is precisely Born's rule.

We simulated the dynamical equation \eqref{eq:Collapse_SDE_final_concise} using Monte Carlo methods to verify this behavior. As shown in Fig.~\ref{collapsfig}A, a single simulation run reveals that the trajectory randomly and irreversibly collapses to a single eigenstate, with the randomness driven by the electromagnetic white noise. In Fig.~\ref{collapsfig}B, we perform multiple simulation runs starting from the same initial state $|\Psi(0)\rangle = \frac{\sqrt{3}}{2}|\uparrow\rangle + \frac{1}{2}|\downarrow\rangle$. As the number of trials increases, the relative frequencies of collapsing to $|\uparrow\rangle$ and $|\downarrow\rangle$ progressively converge to the squared moduli of the initial state's projections onto these respective eigenstates---demonstrating Born's rule. Expanding on this, Fig.~\ref{collapsfig}C presents Monte Carlo simulations for a variety of initial states. The results demonstrate that as the number of simulations grows, the collapse probabilities consistently approach the theoretical Born line, providing numerical validation for the derived statistics.

\textbf{3) Finite collapse timescale and noise anisotropy.} Unlike the instantaneous collapse postulated by the Copenhagen interpretation, this stochastic dynamics predicts that measurement is a rapid but finite physical process, characterized by a mean first-passage time. Consider a two-level system spanned by the pointer basis $\{|\varphi_1\rangle, |\varphi_2\rangle\}$. Defining the projection probability $p_1(t) \equiv |a_1(t)|^2 = 1 - |a_2(t)|^2$ with an initial value $p_{\text{ini}} = p_1(0)$, the vector SDE, Eq.~\eqref{eq:Collapse_SDE_final_concise}, reduces to a one-dimensional diffusion process:
\begin{equation}
	\mathrm{d}p_1 = \sigma_{\text{eff}}\, p_1(1-p_1) \mathrm{d}W_t.
\label{eq:1D_diffusion}
\end{equation}
Here, we consider directional noise, where $\mathrm{d}\vec{W} = \vec{e}_W\mathrm{d}W_t$. The unit vector $\vec{e}_W$ denotes the polarization direction of the local electromagnetic noise. The effective diffusion rates are given by $\sigma_{\text{eff}} = 2 \sigma_E \left| \vec{N}_{1,2} \cdot \vec{e}_W \right|$ and $\sigma_{\text{eff}} = 2 Z_{\text{cr}} \sigma_B \left| \vec{M}_{1,2} \cdot \vec{e}_W \right|$, for purely orbital (electric) and purely spin (magnetic) collapse channels, respectively.

The collapse time $\tau$, defined as the mean first-passage time, is governed by the Pontryagin backward equation  \cite{risken1996fokker, gardiner1985handbook}:
\begin{equation}
	\frac{1}{2} \sigma_{\text{eff}}^2\,p_{\text{ini}}^2\,(1-p_{\text{ini}})^2 \frac{\mathrm{d}^2 \tau}{\mathrm{d}p_{\text{ini}}^2} = -1.
\label{eq:pontryagin_backward}
\end{equation}
A subtlety arises here: because $p_{\text{ini}}^2\,(1-p_{\text{ini}})^2$ vanishes at the ideal vertices $p_{\text{ini}}=0$ and $p_{\text{ini}}=1$, the backward equation is singular at the exact absorbing boundaries. Equivalently, the mean first-passage problem to the perfectly collapsed states is not operationally well posed at infinite precision.

In practice, however, a measurement never resolves an infinitely exact pure state. It is therefore natural to introduce a small resolution threshold $\epsilon\ll1$ and define the collapse as complete once the trajectory first enters either interval $[0,\epsilon]$ or $[1-\epsilon,1]$. The mean collapse time is then determined by the regularized boundary-value problem $\tau(\epsilon)=0,
	\tau(1-\epsilon)=0$. Under these boundary conditions, Eq.~\eqref{eq:pontryagin_backward} yields
\begin{equation*}
	\tau(p_{\text{ini}})=\frac{2}{\sigma_{\rm eff}^2}\left[(2p_{\text{ini}}-1)\ln\frac{1-p_{\text{ini}}}{p_{\text{ini}}}-(2\epsilon-1)\ln\frac{1-\epsilon}{\epsilon}\right].
\end{equation*}
For the maximally unbiased initial state $p_0=\tfrac12$, this reduces to
\begin{equation}
	\max(\tau)=\tau\left(\frac{1}{2}\right)=\frac{2}{\sigma_{\rm eff}^2}
	(1-2\epsilon)\ln\frac{1-\epsilon}{\epsilon}.
\label{eq:tau_half_epsilon}
\end{equation}
Choosing, for example, $\epsilon=0.005$ corresponds to the practical criterion that the collapse is regarded as complete once one outcome probability reaches $0.995$, i.e. once the state is within $0.5\%$ of a pointer vertex. In that case,
\begin{equation}
	\tau_{\rm clps}=\max(\tau)=
	\tau\left(\frac{1}{2}\right)
	\approx
	10.48\,\sigma_{\rm eff}^{-2}.
\label{eq:tau_half_restate2}
\end{equation}
To establish the practical operational limits of this process, we evaluate the scenario where the environmental noise is optimally aligned with the system's differential susceptibility vector ($\vec{e}_W \parallel \vec{N}_{1,2}$ or $\vec{e}_W \parallel \vec{M}_{1,2}$), which maximizes the projection such that $|\vec{V} \cdot \vec{e}_W|^2 \to |\vec{V}|^2$. Under this maximal-coupling condition, substituting the microscopic expressions for $\vec{N}_{1,2}$ and $\vec{M}_{1,2}$ yields the explicit, fastest characteristic collapse timescales for the electric-field (orbital) and magnetic-field (spin) channels:
\begin{equation}
	\tau_{\text{clps}}^E \approx \frac{1.067\pi m_0^3c^4}{(q\sigma_{E})^2 E_{\text{gap}}}, \qquad\tau_{\text{clps}}^{B}\approx\frac{9.72 \alpha_{\text{EM}}^3 m_0^3c^2}{\pi \sigma_B^2 q^2 k_B T}.
\label{eq:Collapse_Time}
\end{equation}
$\alpha_{\text{EM}}$ here is the fine structure constant. To validate these theoretical predictions, we calculate $\tau_{\mathrm{clps}}$ for two representative solid-state systems. For GaAs quantum dots (orbital channel)  \cite{pettaJR2005, bluhm2011dephasing}, using typical parameters $\sigma_E \approx 2.5~\text{kV}/(\text{m}\sqrt{\text{Hz}})$, $E_{\text{gap}} \approx 4~\text{meV}$, and $m_{\text{eff}} \approx 0.067m_e$, Eq.~\eqref{eq:Collapse_Time} yields a theoretical collapse time of $\tau_{\mathrm{clps}}^{E}\approx 60~\text{ns}$. This consistent at the order-of-magnitude level with observed orbital dephasing times of $\sim 10~\text{ns}$  \cite{pettaJR2005}. For the spin channel in diamond NV centers  \cite{mclellan2018nitrogen}, assuming $\sigma_B \approx 0.1~\mu\text{T}/\sqrt{\text{Hz}}$, the predicted timescale scales inversely with temperature. At $T=4~\text{K}$, the model gives $\tau_{\mathrm{clps}}^{B}\approx 5.64~\mu\text{s}$, consistent with measured spin-dephasing times of $\sim 3~\mu\text{s}$. At room temperature ($T=300~\text{K}$), the timescale rapidly decreases to $\tau_{\mathrm{clps}}^{B}\approx 76.76~\text{ns}$, aligning with the nanosecond-scale dephasing observed under thermal conditions. 

The numerical discrepancies may be related to the fact that factors such as environmental noise anisotropy, limited measurement bandwidth, and particle density are not taken into account. In addition, we need to mention that the collapse time is not exactly the same as the decoherence time; They differ by one factor (we can see the difference in Eq.\eqref{EQ:dephasing_rate_main} in Section\ref{sec:decoherence}), which is one of the reasons why our theoretical predictions are little large.

\subsection{Mass scaling and the macroscopic limit}
Equation~\eqref{eq:Collapse_Time} predicts a single-particle scaling $\tau_{\mathrm{clps}}^{E,B}\propto m_0^{3}$, implying that for an isolated particle subject to a fixed noise strength, the collapse time increases with its rest mass. This scaling initially appears to contradict the empirical observation that macroscopic objects decohere rapidly. The apparent contradiction arises from comparing two physically distinct concepts of mass. The $\tau_{\mathrm{clps}}$ scaling refers to the intrinsic rest mass of a single fundamental carrier evolving under a single stochastic channel. In contrast, macroscopic decoherence involves a composite system with a large number of constituents, internal degrees of freedom, and collective couplings to the environment. In such systems, the overall decoherence rate scales with the particle number and the multitude of environmental interaction channels, leading to an effective enhancement that substantially shortens coherence times. Therefore, the rapid decoherence of macroscopic bodies does not contradict the single-particle scaling. Rather, it is a consequence of the many-body amplification of the same underlying stochastic mechanism.

For the spin channel, the dependence $\tau_{\mathrm{clps}}^{B} \propto m_0^3$ arises directly from the inverse scaling of the magnetic moment, $\mu \approx q\hbar/2m_0$. Heavier particles possess smaller magnetic moments and thus couple more weakly to magnetic noise, suppressing the diffusion on the probability simplex. This mechanism explains the well-established experimental fact that nuclear spins (heavy) exhibit coherence times orders of magnitude longer than electron spins (light) in comparable solid-state environments. For instance, $^{31}$P nuclear spins in silicon can retain coherence for over 30 seconds  \cite{muhonen2014storing}, and nuclear spins near diamond NV centers demonstrate second-scale coherence at room temperature  \cite{childress2006coherent, maurer2012room}, far exceeding their electronic counterparts. These results confirm that for single degrees of freedom, larger mass (and thus smaller $\mu$) indeed offers protection against magnetic decoherence.

For the orbital channel, as demonstrated in matter-wave interferometry, large molecules can sustain coherence if environmental scattering is suppressed \cite{arndt1999wave, hornberger2012colloquium}. For macroscopic systems, the decoherence rate grows with the number of environmental coupling channels (typically proportional to system size), leading to an extremely rapid suppression of off-diagonal coherence terms in the reduced density matrix. This dynamical amplification of environmental information flow makes classical behavior effectively unavoidable. This collective scaling drives $\tau_{\mathrm{clps}} \to 0$ despite the protective inertial factor of the individual constituents.

\section{Relativistic path-integral origin of quantum decoherence}
\label{sec:decoherence}
\begin{figure*}[htpb] 
    \centering 
    \includegraphics[width=0.98\textwidth]{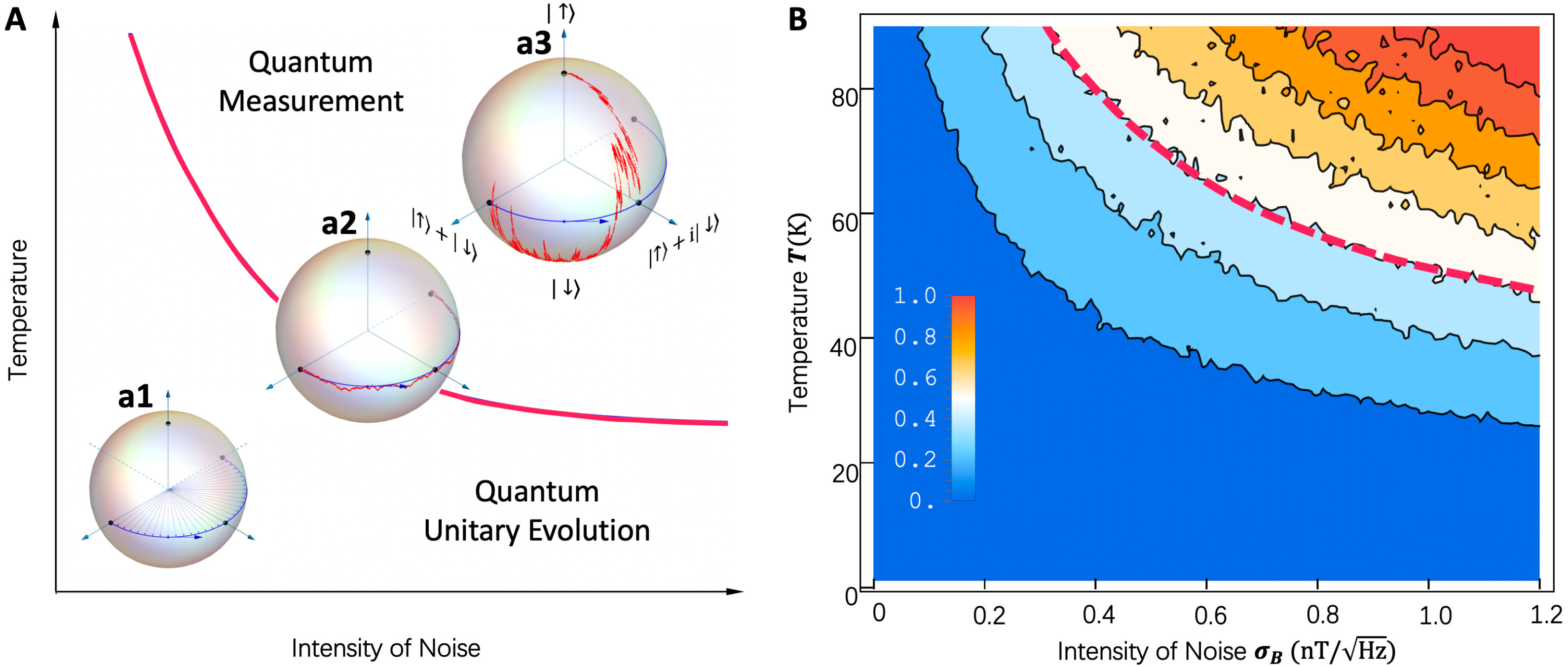} 
	\caption{\textbf{Crossover from unitary evolution to measurement-like collapse.} \textbf{A) Schematic Bloch-sphere snapshots at three noise levels (a1~$\to$~a3)}. The thick red curve is a visual guide to the transition between the unitary and measurement-dominated regimes. In \textbf{a1} (negligible noise or low temperature) the state follows a smooth, unitary equatorial trajectory (blue); under a $\pi$-pulse of duration $\tau_\pi = {2 m \pi}/{(g q B)}$ it reaches the target state $|\uparrow\rangle-|\downarrow\rangle$. In \textbf{a2} (moderate noise or moderate temperature) the stochastic path on the sphere (red) deviates from the unitary arc but, within one $\pi$-pulse, still ends near the same target---this is the critical region: increasing the temperature $T$ or the noise amplitude pushes the dynamics across the boundary. In \textbf{a3} (strong noise or high temperature) the trajectory becomes erratic (red) and, within a $\pi$-pulse, the state randomly collapses to $|\uparrow\rangle$ or $|\downarrow\rangle$. \textbf{B) Monte-Carlo phase map of the collapse weight $W_M$ (see Methods) as a function of magnetic-noise intensity $\sigma_B$ ($\mathrm{nT}/\sqrt{\mathrm{Hz}}$) and temperature $T (\mathrm{K})$}. The map is computed on a $300\times300$ grid in $(\sigma_B,T)$; at each grid point $W_M$ is estimated by averaging over 2000 stochastic trajectories of the collapse SDE. Colors run from blue ($W_M \approx 0$, near-unitary) to red ($W_M \approx 1$, measurement limit), with black iso-contours. The dashed red line overlays the same transition guide shown in panel~\textbf{A} and closely follows the $W_M \approx 0.5$ contour. Together, panels \textbf{A} and \textbf{B} show that unitary evolution and quantum measurement arise as two limits of a single mechanism.}
    \label{fig:unification} 
\end{figure*}

The preceding analysis established trajectory-level collapse dynamics driven by non-differentiable relativistic electromagnetic fluctuations. During this process, the coherent dynamical phase generated by $\hat H_0$ over an interval $\mathrm dt$ is higher order than the stochastic increment $\mathrm d\vec W$. Consequently, under strong noise or on very short timescales, coherent evolution is effectively masked by noise-induced fluctuations. For moderate noise intensities or over longer evolution times, however, the coherent drift cannot be neglected. In that regime, unitary evolution driven by smooth potentials and collapse driven by noisy potentials compete, yet both are governed by the same evolution operator $\hat U$. For a fixed operation time, weak noise leads to unitary-dominated dynamics and predictable evolution outcomes, whereas strong noise leads to measurement-dominated dynamics and randomized results. Our theory predicts the boundary between these regimes (Fig.~\ref{fig:unification}).

This unified dynamical mechanism bridges microscopic state reduction with macroscopic open-system dynamics, yielding the Gorini-Kossakowski-Sudarshan-Lindblad (GKSL) master equation without invoking the traditional Born-Markov approximations. Therefore, it provides two complementary physical descriptions: (i) The conditional (selective) evolution, conditioned on a specific environmental noise (measurement) record, generates stochastic quantum trajectories corresponding to single-run collapse events; and (ii) The unconditional (non-selective) evolution, obtained by discarding this record and averaging over all noise realizations, yields a deterministic evolution of the ensemble-averaged state, manifesting as continuous decoherence. This rigorous mathematical correspondence show that measurement-like collapse and decoherence arise as two complementary statistical descriptions of the same noise-activated relativistic dynamics.

\subsection{Unified evolution equation and the unitary--measurement boundary}
In the It\^o white-noise limit, the infinitesimal propagator takes the form (see Appendix \ref{app:stochastic_timeslice_U})
\begin{equation}
\label{eq:shortprop_main}
	\begin{aligned}
		\hat U(t,t+\mathrm dt)=&\mathbf 1-\frac{\mathrm i}{\hbar}\hat H_0\mathrm dt+\sum_\mu \hat J_\mu\mathrm dW_\mu \\
		&-\frac{1}{2}\sum_{\mu,\nu}G_{\mu\nu}\hat J_\nu\hat J_\mu\mathrm dt+o(\mathrm dt).
	\end{aligned}
\end{equation}
Here $\mathrm dW_\mu$ are Wiener increments satisfying $\mathbb E[\mathrm dW_\mu]=0$ and $\mathrm dW_\mu\mathrm dW_\nu=G_{\mu\nu}\mathrm dt$, where $G=[G_{\mu\nu}]\succeq 0$ is a constant covariance matrix  \cite{Gardiner2009,Oksendal2013}. The operators $\hat J_\mu$ are Hermitian noise couplings (see Eq.~\eqref{eq:new_definition}).

The quadratic term $\hat{J}_{\nu}\hat{J}_{\mu}$ in Eq.~\eqref{eq:shortprop_main} is necessary. Because $\mathrm dW_\mu=O(\sqrt{\mathrm dt})$ and $\mathrm dW_\mu\mathrm dW_\nu=O(\mathrm dt)$, It\^o order counting forces the $O(\mathrm dt)$ correction. Without it, the ensemble-averaged map would fail to reproduce a completely positive, trace-preserving (CPTP) generator \cite{Gardiner2009,Oksendal2013,Breuer2002}.

Figure~\ref{fig:unification} illustrates the evolution of a state $|\Psi(t)\rangle=\hat U(t,0)|\Psi(0)\rangle$ for a two-level system. As visualized on the Bloch sphere trajectories in Fig.~\ref{fig:unification}A, weak noise permits near-unitary evolution toward a coherent target state, whereas strong noise overrides the dynamics, forcing erratic, measurement-dominated outcomes. 

To quantify this crossover, we define the collapse weight $W_M = \big|p_1(t)-p_2(t)\big|\in[0,1]$, where $p_1$ and $p_2$ are the occupation probabilities of the pointer states. For a coherent equatorial superposition, $W_M=0$ (pure unitarity), while for absolute collapse to a definitive pointer state, $W_M=1$ (pure measurement). Our Monte Carlo simulations map this transition (Fig.~\ref{fig:unification}B), revealing a distinct boundary curve that separates the unitary and measurement regimes. This phase boundary is determined analytically by the condition $\tau_{\mathrm{clps}}=\tau_\pi$, where $\tau_\pi$ is the unitary gate time, confirming the theoretical scaling.

\subsection{Conditional dynamics and the innovation superoperator}
Given a normalized conditional state $\rho(t)$, the evolution over one time slice is
\begin{equation}
\label{eq:tilderho_evolution_main}
	\tilde\rho(t+\mathrm dt)=\hat U(t,t+\mathrm dt)\rho(t)\hat U^\dagger(t,t+\mathrm dt).
\end{equation}
We use $\tilde\rho$ to emphasize that the evolution is linear but not trace-preserving along a single realization because $\hat U$ is non-unitary in the presence of stochastic action. Consequently, further normalization is required:
\begin{equation}
\label{eq:rho_norm_main}
	\rho(t+\mathrm dt)=\frac{\tilde\rho(t+\mathrm dt)}{\Tr\big[\tilde\rho(t+\mathrm dt)\big]}.
\end{equation}
Substituting Eq.~\eqref{eq:shortprop_main} into Eq.~\eqref{eq:rho_norm_main} and retaining terms up to $O(\mathrm dt)$ yields a normalized diffusive stochastic master equation (SME) of the standard form
\begin{equation}
\label{eq:SME_methods}
	\begin{aligned}
		\mathrm d\rho &=-\frac{\mathrm i}{\hbar}[\hat H_0,\rho]\mathrm dt \\
		&+\sum_{\mu,\nu}G_{\mu\nu}\mathcal{D}[\hat{J}_{\mu},\hat{J}_{\nu}][\rho]\mathrm dt+\sum_\mu \mathcal H[\hat J_\mu]\rho\mathrm dW_\mu^{\mathrm{in}}.
	\end{aligned}
\end{equation}
In this derivation we use $\hat J_\mu^\dagger=\hat J_\mu$. The superoperators are \cite{WisemanMilburn2009}
\begin{equation}
\label{eq:superops_defs}
	\begin{aligned}
		&\mathcal{D}[\hat{J}_{\mu},\hat{J}_{\nu}][\rho]=\hat J_\mu\rho\hat J_\nu-\frac12\{\hat J_\nu\hat J_\mu,\rho\}, \\ 
		&\mathcal H[\hat{J}_\mu]\rho=\hat{J}_\mu\rho+\rho\hat{J}_{\mu}-2\langle \hat{J_\mu}\rangle\rho,
	\end{aligned}
\end{equation}
where $\{A,B\}=AB+BA$ is the anticommutator and $\langle \hat J\rangle\equiv \Tr(\hat J\rho)$. $\mathcal{D}[\hat{J}_{\mu},\hat{J}_{\nu}]$ is the dissipator term that survives ensemble averaging, while $\mathcal H[\hat{J}_{\mu}]$ is the nonlinear innovation superoperator that subtracts the conditional expectation to preserve normalization along each trajectory.

After the normalization step \eqref{eq:rho_norm_main}, the natural increments driving the conditional dynamics are the innovation Wiener increments  \cite{WisemanMilburn2009,JacobsSteck2006,Bouten2007}
\begin{equation}
\label{eq:innovation_main}
	\mathrm dW_\mu^{\mathrm{in}}=\mathrm dW_\mu-2\sum_\nu G_{\mu\nu}\langle \hat J_\nu\rangle\mathrm dt.
\end{equation}
This is a Girsanov-type shift induced by the normalization step: trajectory weights are continuously reweighted, and under the physical conditional measure the innovation has zero conditional mean, $\mathbb E[\mathrm dW_\mu^{\rm in}\mid \mathcal F_t]=0$, while preserving the same quadratic variation, $\mathrm dW_\mu^{\rm in}\mathrm dW_\nu^{\rm in}=G_{\mu\nu}\,\mathrm dt$. Thus the covariance is unchanged, but the state-dependent drift carried by the raw noise is absorbed into the innovation process.

We note that the noise term appearing in Eq.~\eqref{eq:collapse_equation_ito_form} is likewise understood as an innovation Wiener process. In the measurement section, however, we focused on the instantaneous evolution and therefore retained only terms of order $\sqrt{\mathrm{d}t}$, neglecting all $\mathrm{d}t$-order contributions. Since the distinction between a raw Wiener increment and its innovation counterpart manifests only through drift corrections at order $\mathrm{d}t$, this truncation leaves the martingale structure of the probability dynamics unaffected. The term with order $\mathrm{d}t$ appears only when deriving the full stochastic master equation at order $\mathrm{d}t$, as done in the present section.

Eq.~\eqref{eq:SME_methods} describes the conditional (record-conditioned) dynamics of the quantum state. For Hermitian diffusive couplings, it preserves purity for pure initial conditions in the ideal-efficiency limit. In the commuting limit $[\hat H_0,\hat J_\mu]=0$, the SME(stochastic master equation) closes on the diagonal populations in the common eigenbasis and reduces to the bounded-martingale probability diffusion derived in Sec.~\ref{sec:measurement}. 

\subsection{Unconditional dynamics and the GKSL master equation}
Discarding the record and averaging over realizations yields the ensemble state $\bar\rho(t)\equiv \mathbb E[\rho(t)]$. Unlike individual conditional trajectories, which require active continuous normalization to compensate for measurement backaction, this ensemble average follows a fundamentally linear, trace-preserving evolution. The ensemble dynamics can be obtained in two equivalent ways. One may average the linear map \eqref{eq:tilderho_evolution_main}, for which all terms linear in $\mathrm dW_\mu$ vanish because $\mathbb E[\mathrm dW_\mu]=0$, while the It\^o contractions $\mathbb E[\mathrm dW_\mu\mathrm dW_\nu]=G_{\mu\nu}\mathrm dt$ generate the dissipator. Equivalently, averaging the normalized SME \eqref{eq:SME_methods} removes the innovation term because $\mathrm dW_\mu^{\mathrm{in}}$ has zero conditional mean. 

In both cases one obtains the time-homogeneous GKSL master equation  \cite{Gorini1976,Lindblad1976,Breuer2002}
\begin{equation}
\label{eq:GKSL}
	\frac{\mathrm d\bar\rho}{\mathrm dt}=-\frac{\mathrm i}{\hbar}[\hat H_0,\bar\rho]+\sum_{\mu,\nu}G_{\mu\nu}\!\left(\hat J_\mu\bar\rho\hat J_\nu-\frac12\{\hat J_\nu\hat J_\mu,\bar\rho\}\right).
\end{equation}
Complete positivity is guaranteed by $G=[G_{\mu\nu}]\succeq 0$, which is the same positivity condition required for the Wiener covariance \cite{Breuer2002,Gorini1976,Lindblad1976}. This result makes the relation between measurement and decoherence connection explicit: the GKSL equation is not introduced phenomenologically, nor obtained by separately imposing a Born--Markov approximation on a reduced system. Rather, single-shot collapse arises from the conditional (record-conditioned) evolution of the same relativistic noise-activated trajectory dynamics. While the conditional SME retains the record-conditioned nonlinear innovation terms responsible for wave-function collapse, the unconditional GKSL master equation captures only their ensemble-averaged effect, appearing as deterministic decoherence.

The relation becomes especially transparent when $\hat H_0$ and all $\hat J_\mu$ commute. In their common eigenbasis,
\begin{equation*}
	\hat H_0|\varphi_j\rangle=E_j|\varphi_j\rangle,\qquad
	\hat J_\mu|\varphi_j\rangle=J_{\mu,j}|\varphi_j\rangle,
\end{equation*}
the diagonal populations satisfy $\frac{\mathrm d}{\mathrm dt}\bar\rho_{jj}=0$,while the off-diagonal terms obey
\begin{equation}
\label{eq:dephasing_rate_main}
	\begin{aligned}
		\frac{\mathrm d}{\mathrm dt}\bar\rho_{jk}&=-\frac{\mathrm i}{\hbar}(E_j-E_k)\bar\rho_{jk}-\Gamma_{jk}\bar\rho_{jk},\\ 
		&\Gamma_{jk}=\frac12\sum_{\mu,\nu}G_{\mu\nu}J_{\mu}^{(j,k)} J_{\nu}^{(j,k)},
	\end{aligned}
\end{equation}
with $J_{\mu}^{(j,k)}\equiv J_{\mu,j}-J_{\mu,k}$. Thus the decoherence rate is directly the quadratic form of the same differential susceptibilities that control the collapse diffusion in Sec.~\ref{sec:measurement}. In particular, for an effectively single-direction (rank-one) diffusive noise channel acting within a two-level quantum system, one finds the identity $\Gamma_{12}={\sigma_{\rm eff}^2}/{8}$, where $\sigma_{\rm eff}$ is the conditional diffusion strength entering the single-trajectory probability dynamics. The characteristic ensemble dephasing time is therefore $T_2=\Gamma_{12}^{-1}=8\,\sigma_{\rm eff}^{-2}$. Importantly, this ensemble timescale exhibits the identical $\sigma_{\rm eff}^{-2}$ scaling as the single-shot collapse time $\tau_{\rm clps}\approx {10.48}\,\sigma_{\rm eff}^{-2}$ obtained from the first-passage analysis in Eq.~\eqref{eq:tau_half_restate2}. The slight difference in the prefactor arises purely from their distinct statistical definitions: $T_2$ represents the $1/e$ exponential decay of an ensemble average, whereas $\tau_{\text{clps}}$ represents the mean time for a single stochastic trajectory to hit a definitive $99\%$ absorption boundary.

This analytical mapping demonstrates, within a single theoretical structure, how the speed of single-shot wave-function collapse and the rate of ensemble decoherence are governed by the identical microscopic relativistic noise couplings, differing only by the mathematical perspective of conditioning versus averaging.

Finally, Eq.~\eqref{eq:GKSL} characterizes the Markovian regime implied by the It\^o white-noise limit. If the underlying relativistic noise has finite correlation time, the conditional trajectory dynamics remain well defined, but the ensemble evolution generally acquires memory kernels or time-dependent coefficients. This transition into the non-Markovian regime and its implications for measurement are analyzed in Sec.~\ref{sec:colored_noise}.

\section{Relativistic path-integral origin of non-Hermitian physics}
\label{sec:nonhermitian}

The preceding sections established that non-differentiable electromagnetic fluctuations induce a local stochastic propagation which, upon averaging over the noise record, yields a completely positive and trace-preserving (CPTP) ensemble evolution. We now extend this exact construction to a composite bipartite system $AB$ to address a fundamental physical question: if the noisy environment acts strictly locally on subsystem $B$, what dynamics are structurally induced on the reduced density matrix of subsystem $A$?

This scenario directly intersects the theoretical domain of ``effective non-Hermitian'' (NH) quantum mechanics. Standard phenomenological NH models routinely postulate an anti-Hermitian operator on $B$, which mathematically generates a deterministic drift on the remote state $A$. If treated as an unconditional physical law, this directly violates the no-signaling principle of special relativity. The derivation below resolves this paradox. We demonstrate that the exact algebraic tensor structure of NH physics arises naturally from our local stochastic relativistic propagator, but it enters exclusively through zero-mean It\^o innovations. Consequently, the unconditional state of $A$ remains invariant under local dynamics on $B$, rigorously preserving relativistic causality while fully recovering the conditioned dissipation characteristic of NH theories.

\subsection{Local stochastic propagator on $B$ and the reduced dynamics on $A$}
Consider a bipartite state $\rho_{AB}(t)$ defined on the Hilbert space $\mathcal H_A \otimes \mathcal H_B$, with dimensions $\dim \mathcal H_A = n$ and $\dim \mathcal H_B = m$. At this point, the infinitesimal time-evolution operator acting on system B is
\begin{equation*}
	\begin{aligned}
		\hat U_B(t,t+\mathrm dt)=&I_B-\frac{\mathrm i}{\hbar}\hat H_B \mathrm dt+\sum_{\mu}\hat{J}_\mu \mathrm dW_\mu \\
		&-\frac{1}{2}\sum_{\mu,\nu}G_{\mu\nu}\hat{J}_\nu \hat{J}_\mu \mathrm dt+o(\mathrm dt).
	\end{aligned}
\end{equation*}
The quadratic It\^o term is dictated by order counting ($\mathrm dW_\mu \mathrm dW_\nu = G_{\mu\nu}\mathrm dt$) and is mathematically required to ensure that the ensemble-averaged evolution remains completely positive and trace-preserving. As a result, the density matrix of this bipartite system evolves into
\begin{equation}
	\tilde\rho_{AB}(t+\mathrm dt)=(I_A \otimes \hat U_B)\rho_{AB}(t)(I_A \otimes \hat U_B^\dagger).
\label{eq:rhoAB_tilde_evolution}
\end{equation}
Expanding Eq.~\eqref{eq:rhoAB_tilde_evolution} rigorously to $\mathcal O(\mathrm dt)$ yields
\begin{equation*}
	\mathrm d\tilde\rho_{AB}=\mathcal{G}_B(\rho_{AB})\mathrm dt+\sum_\mu\{(I_A\otimes \hat{J}_\mu),\rho_{AB}\}\mathrm dW_\mu,
\end{equation*}
where $\mathcal{G}_B$ is the standard GKSL generator acting strictly locally on subsystem $B$ \cite{Gorini1976,Lindblad1976,Breuer2002}. We evaluate the induced dynamics on the remote subsystem by tracing out $B$, defining $\tilde\rho_A := \Tr_B(\tilde\rho_{AB})$. Because the GKSL generator is inherently trace-preserving, $\Tr_B\!\left[\mathcal{G}_B(X)\right] = 0$ for any bipartite operator $X$. Therefore, the $\mathrm dt$ drift term vanishes identically, leaving the evolution of $\rho_A$ purely stochastic:
\begin{equation}
	\mathrm d\tilde\rho_A=\sum_\mu\Tr_B\Big(\{(I_A\otimes \hat{J}_\mu),\rho_{AB}\}\Big)\mathrm dW_\mu.
\label{eq:drhoA_basic}
\end{equation}
Equation~\eqref{eq:drhoA_basic} inherently satisfies the no-signaling constraint. The physical influence of the local noise on $B$ propagates to $A$ strictly through the Wiener innovations $\mathrm dW_\mu$. Because these stochastic increments possess zero expectation ($\mathbb E[\mathrm dW_\mu]=0$), the unconditional reduced state $\rho_A(t)=\Tr_B\big(\rho_{AB}(t)\big)$ remains strictly invariant under the local measurement dynamics on $B$:
\begin{equation}
\mathbb E[\mathrm d\tilde\rho_A]=0
\quad \Rightarrow \quad
\mathrm d\rho_A = 0.
\label{eq:nosignaling_statement}
\end{equation}
This mathematically proves that local relativistic noise cannot be leveraged to transmit deterministic signals to a remote subsystem, conforming to quantum kinematic constraints \cite{NielsenChuang2000,Breuer2002}.

\subsection{Hilbert--Schmidt decomposition and the three-term tensor structure}
To explicitly evaluate the partial trace in Eq.~\eqref{eq:drhoA_basic}, we employ a canonical Hilbert--Schmidt decomposition. Let $\{F_\alpha\}_{\alpha=0}^{n^2-1}$ and $\{G_\beta\}_{\beta=0}^{m^2-1}$ form orthonormal Hermitian operator bases for $\mathcal H_A$ and $\mathcal H_B$, normalized such that $F_0=I_A/\sqrt n$, $G_0=I_B/\sqrt m$, and $\Tr(F_\alpha)=\Tr(G_\beta)=0$ for $\alpha,\beta\ge 1$. The generic bipartite state is expressed as
\begin{equation}
	\rho_{AB}=\rho_A\otimes\frac{I_B}{m}+\frac{I_A}{n}\otimes\rho_B-\frac{I_A}{n}\otimes\frac{I_B}{m}+T,
\label{eq:rhoAB_decomp_main_nonH}
\end{equation}
where
\begin{equation*}
	T=\sum_{\alpha=1}^{n^2-1}\sum_{\beta=1}^{m^2-1}t_{\alpha\beta}F_\alpha\otimes G_\beta
\end{equation*}
is the core correlation tensor satisfying $\Tr_A(T)=\Tr_B(T)=0$. Defining the scalar overlaps
\begin{equation*}
	s_\mu := \Tr(\frac{\hat{J}_\mu}{m}),\ \ \ell_\mu:=\Tr(\hat{J}_\mu\rho_B),\ \ c_{\mu\beta}:=\Tr(\hat{J}_\mu G_\beta),
\end{equation*}
substituting Eq.~\eqref{eq:rhoAB_decomp_main_nonH} into Eq.~\eqref{eq:drhoA_basic} isolates the induced stochastic evolution into three structural contributions:
\begin{equation}
	\begin{aligned}
		\mathrm d\tilde\rho_A= \sum_{\mu}\Bigg[&\Big(\rho_A-\frac{I_A}{n}\Big){s}_\mu+\frac{I_A}{n}{\ell}_\mu \\
		&+\sum_{\alpha=1}^{n^2-1}\sum_{\beta=1}^{m^2-1}t_{\alpha\beta}F_\alpha c_{\mu\beta}\Bigg]\mathrm dW_\mu.
	\end{aligned}
\label{eq:rhoA_ours}
\end{equation}
Each contribution has a distinct operational origin, fixed purely by Hilbert--Schmidt algebra and the identity/traceless splitting on $B$.

\textbf{1) Trace contribution of the local channel.}
The coefficient $s_\mu=\Tr(\hat J_\mu/m)$ extracts the identity component of the local coupling on $B$. When $s_\mu\neq 0$, the local propagator on $B$ contains a part proportional to $I_B$. Through the partial trace, this produces a stochastic reweighting of the remote state proportional to $\rho_A-I_A/n$. This term vanishes for a traceless monitoring channel (the typical case for dephasing measurements, where $\Tr \hat J_\mu=0$), and it vanishes when $\rho_A$ is maximally mixed. Because it multiplies $\mathrm dW_\mu$, it generates no deterministic relaxation of the unconditional state. It represents a record-dependent fluctuation that disappears upon averaging, conforming to the no-signaling constraint.

\textbf{2) Local measurement backaction from the state of $B$.}
The scalar $\ell_\mu=\Tr(\hat J_\mu\rho_B)$ is the instantaneous expectation value of the monitored observable on $B$. It contributes exclusively through the identity operator on $A$. Consequently, it does not alter the traceless part of $\tilde\rho_A$ (it neither rotates nor polarizes $A$); instead, it modulates the overall unnormalized weight of the reduced state of the remote subsystem along the trajectory. This term persists even for separable product states. It reflects a likelihood reweighting conditioned on the local measurement record of $B$, rather than a nonlocal physical influence.

\textbf{3) Correlation component and entanglement-enabled steering.}
The third term is proportional to the correlation tensor $t_{\alpha\beta}$. It vanishes identically for product states ($T=0$). This contribution survives only when the bipartite system $AB$ carries correlations, enabling the reduced evolution on $A$ to be modulated by the local noise on $B$. This is the only term that alters the direction of $\tilde\rho_A$ (its traceless component) in a record-dependent manner, thereby realizing the coordinated quantum steering effect. Conditioning on the local innovations on $B$ selects a specific branch of $\rho_{AB}$, and the remote subsystem on $A$ evolutions coherently with that selection. Because this evolution enters exclusively through the zero-mean innovation $\mathrm dW_\mu$, it averages to zero. Thus, the unconditional state $\rho_A$ remains invariant under the local CPTP dynamics on $B$.

\subsection{Resolution of the non-Hermitian signaling paradox}
In phenomenological quantum-trajectory theory, it is standard practice to introduce an effective non-Hermitian Hamiltonian on $B$, defined as
\begin{equation*}
	\hat H_{\mathrm{eff}}=\hat H_B^{(\mathrm{Re})}-\mathrm i\hat\Gamma_B,\qquad\hat\Gamma_B=\hat\Gamma_B^\dagger\ge 0.
\end{equation*}
This operator is designed to generate a continuous ``no-jump'' decay  \cite{Dalibard1992,Dum1992,PlenioKnight1998}. The postulated joint linear evolution is
\begin{equation}
	\frac{\mathrm d\rho_{AB}}{\mathrm dt}=-\frac{\mathrm i}{\hbar}\Big[(I_A\otimes \hat H_{\mathrm{eff}})\rho_{AB}-\rho_{AB}(I_A\otimes \hat H_{\mathrm{eff}}^\dagger)\Big].
\label{eq:NH_postulate}
\end{equation}
Tracing over $B$ eliminates the commutator, yielding an apparent deterministic drift on the remote state:
\begin{equation*}
	\dot\rho_A=-\frac{1}{\hbar}\Tr_B\Big(\big\{I_A\otimes \hat\Gamma_B,\ \rho_{AB}\big\}\Big).
\label{eq:rhoA_NH_start}
\end{equation*}
Using the same decomposition in Eq.~\eqref{eq:rhoAB_decomp_main_nonH} and defining
\begin{equation*} 
	h:=\frac{1}{m}\Tr(\hat\Gamma_B),\ \ \lambda:=\Tr(\rho_B\hat\Gamma_B),\ \ d_\beta:=\Tr(G_\beta\hat\Gamma_B), 
\label{eq:Gamma_overlaps} 
\end{equation*} 
one obtains 
\begin{equation} 
	\begin{aligned} 
		\mathrm{d}\rho_A=-\frac{2}{\hbar}\Bigg[&\Big(\rho_A-\frac{I_A}{n}\Big)h+\frac{I_A}{n}\lambda \\ 
		&+\sum_{\alpha=1}^{n^2-1}\sum_{\beta=1}^{m^2-1}t_{\alpha\beta}F_\alpha d_\beta\Bigg]\mathrm{d}t. 
	\end{aligned}
\label{eq:rhoA_NH}
\end{equation}
A direct algebraic comparison between Eq.~\eqref{eq:rhoA_ours} and Eq.~\eqref{eq:rhoA_NH} reveals an exact structural isomorphism: both the stochastic path-integral evolution and the non-Hermitian effective drift generate the identical three-term tensor decomposition, $(\rho_A-\tfrac{I_A}{n})$, $\tfrac{I_A}{n}$, $\sum_{\alpha,\beta}t_{\alpha\beta}F_\alpha(\cdot)_\beta$. No additional dynamical assumption is involved; the coincidence follows purely from Hilbert--Schmidt operator algebra. 

The decisive distinction is not algebraic but stochastic order. In the non-Hermitian description the entire tensor structure multiplies the deterministic increment $\mathrm dt$. If interpreted as a fundamental, unconditional dynamical law, Eq.~\eqref{eq:rhoA_NH} produces a systematic drift of the reduced state $\rho_A$. Since the drift depends on $\rho_{AB}$, it would generically allow a local operation on $B$ to alter the ensemble state of $A$, contradicting the operational no-signaling principle that follows from local CPTP dynamics. 

By contrast, in the relativistic stochastic construction the same tensor structure multiplies the Wiener increment $\mathrm dW_\mu$. It therefore enters exclusively at order $\sqrt{\mathrm dt}$ and has zero unconditional mean: $\mathbb E[\mathrm dW_\mu]=0$. All remote evolutions are record-dependent innovations and vanish under ensemble averaging. Thus the ensemble-averaged reduced state $\rho_A$ remains invariant under local noise on $B$, while conditioned trajectories exhibit coordinated collapse. 

In this sense, the algebra underlying non-Hermitian drift is not incorrect, but its physical status depends entirely on stochastic order. When treated at order $\mathrm dt$ it appears as a genuine dissipative force; when derived from a relativistic local noise source it appears at order $\mathrm dW$ and becomes a conditional, zero-mean innovation. The apparent contradiction with causality arises only when a conditional generator is misidentified as an unconditional dynamical law. 

The mechanism derived above is not restricted to the single-particle case. When extended to interacting many-body systems, the same structure leads to an effective non-Hermitian description at the macroscopic level. In particular, the operator mismatch generated by non-smooth gauge increments produces directional amplification and attenuation channels in the collective dynamics. These channels naturally reproduce phenomena that are commonly described within non-Hermitian frameworks, including boundary-localized spectral flows and the non-Hermitian skin effect.  A detailed discussion of these topics will be presented in our future work.

\section{Colored noise and directed collapse}
\label{sec:colored_noise}
The derivation of the measurement axioms in Sec.~\ref{sec:measurement} was obtained in the ideal It\^o white-noise limit, where the conditional probability flow on the simplex is an exact bounded martingale. Real electromagnetic environments, however, are neither infinitely broadband nor memoryless: they exhibit finite correlation times, spectral structure, and often pronounced low-frequency components. When our covariant path-integral dynamics is evaluated beyond the white-noise idealization, the collapse law does not disappear; rather, it separates into two physically distinct regimes. In short-memory colored noise the collapse structure is preserved and only the timescale is renormalized, while in long-memory colored noise the martingale property can fail and the absorption odds themselves become steerable.

\subsection{Collapse dynamics properties under short-memory and long-memory noise}

Let the environmental fluctuations be stationary with autocorrelation $C(\tau)$ and a finite correlation time $t_{\mathrm{cor}}$. When the noise memory is short compared with the relevant system timescale (collapse time or gate time), $t_{\mathrm{cor}}\ll \tau_{\mathrm{sys}}$, a standard Markov-embedding/coarse-graining argument reduces the colored driving to an effective diffusion strength set by the integrated spectral weight,
\begin{equation}
\label{eq:Deff_def}
	D_{\mathrm{eff}}:=\int_{0}^{\infty}C(\tau)\,\mathrm d\tau,
\end{equation}
so that the white-noise intensity appearing in the bounded-martingale diffusion is replaced by $\sigma^2\to D_{\mathrm{eff}}$ \cite{Sancho1982,Gardiner2009,Oksendal2013}. In this regime the absorbing boundaries remain the vertices of the probability simplex, and the absorption probabilities remain fixed by the initial weights (Born's rule), because the effective coarse-grained dynamics retains the driftless martingale structure to leading order in $t_{\mathrm{cor}}/\tau_{\mathrm{sys}}$.

\begin{figure*}[htpb] 
    \centering 
    \includegraphics[width=0.98\textwidth]{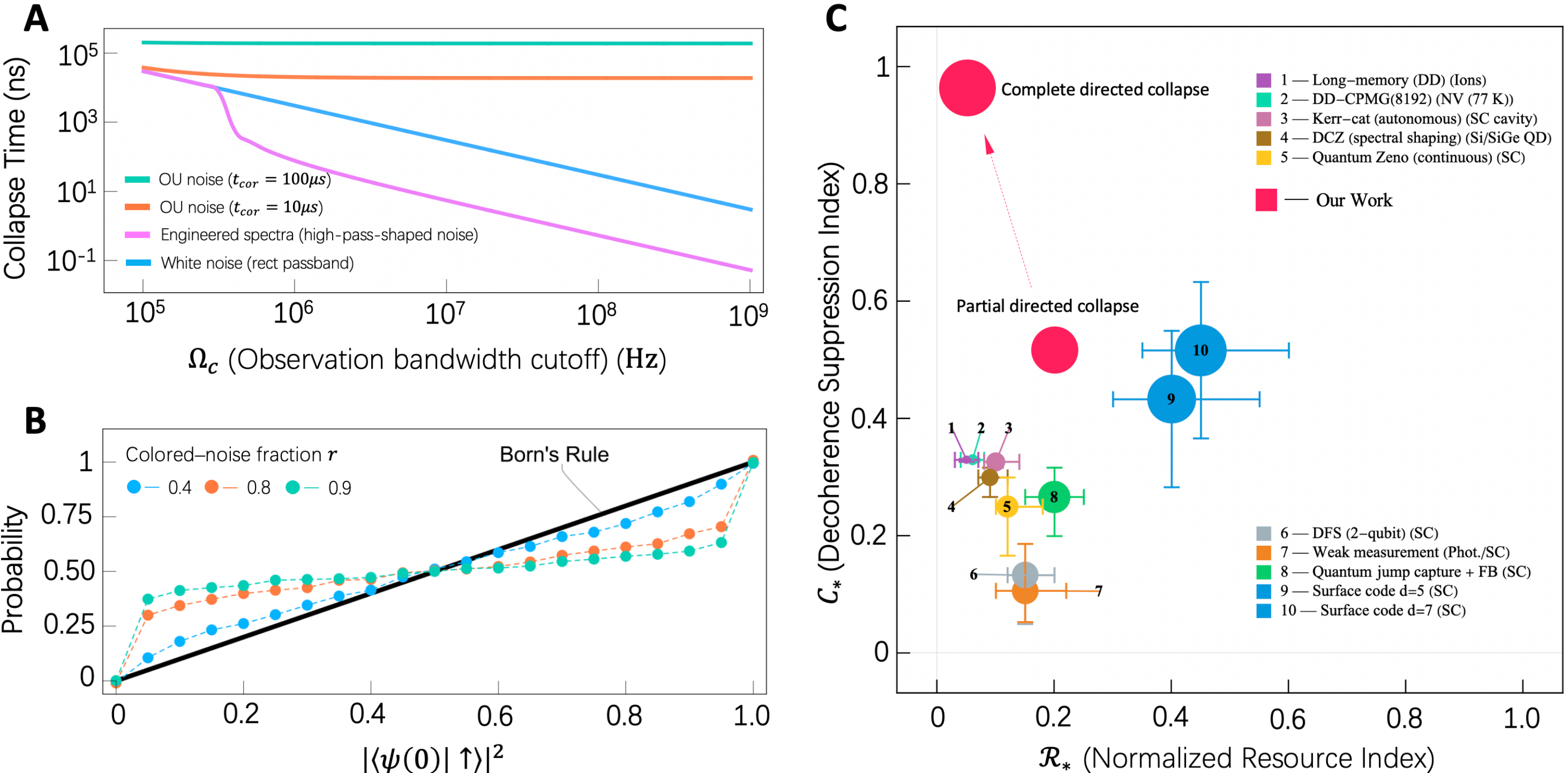} 
	\caption{\textbf{Colored noise steers quantum collapse and advances decoherence reduction technologies.}
	\textbf{(A) Collapse time $\tau_{\mathrm{clps}}$ vs. bandwidth cutoff $\Omega_c$.} Compared to the white-noise baseline (blue), Ornstein-Uhlenbeck noise (orange/green) extends $\tau_{\mathrm{clps}}$ via spectral narrowing, while high-pass shaping (magenta) accelerates collapse by enhancing effective coupling.
	\textbf{(B) Born-line deviations.} Absorption probability $P(\uparrow)$ vs. initial weight $p_0$. While white noise follows the Born rule (black diagonal), increasing the $1/f$ spectral fraction $r$ ($r=S_{\mathrm{1/f}}(0)/[S_{\mathrm{1/f}}(0)+S_{\mathrm{white}}(0)]$) introduces memory-driven drifts that bias outcomes, allowing for controlled deviations.
	\textbf{(C) Resource-effectiveness map.} Benchmarking technical effectiveness ($C_*$,  y-axis; the larger the value, the stronger the suppression of decoherence) against resource cost ($\mathcal{R}_*$, , x-axis; the larger the value, the more qubits, control, or overhead a method requires). The ideal technology would occupy the top-left corner. Formal definitions of the indices are provided in the Methods. Bubble area is proportional to the resource cost, and error bars show the reported performance ranges for established methods (numbered) and our directed-collapse approach (red). The map highlights that our partial-collapse technique is already competitive with existing mid-resource methods, while the complete-collapse variant promises access to a new, highly effective, low-cost regime.
}
\label{fig:colored_noise}
\end{figure*}

What changes is the speed of the diffusion toward the absorbing boundaries. Since the characteristic collapse time scales inversely with the diffusion strength, one has generically
\begin{equation}
\tau_{\mathrm{clps}}\ \propto\ D_{\mathrm{eff}}^{-1},
\end{equation}
with the proportionality constant set by the same first-passage criterion used in Sec.~\ref{sec:measurement}. This provides an immediate control knob. By shaping the noise bandwidth (thereby tuning the transmitted spectral weight entering $D_{\mathrm{eff}}$) and by exploiting the spatial/polarization anisotropy of the fluctuations---which changes the projection of the noise onto the susceptibility-difference directions that enter $\sigma_{\mathrm{eff}}$---one can accelerate collapse for fast reset, or slow it to extend coherence retention, without altering the eventual Born statistics (Fig.~\ref{fig:colored_noise}A).

The situation changes when the noise has substantial long memory, as in spectra with strong low-frequency weight (e.g.\ pronounced $1/f$ components). In that case $t_{\mathrm{cor}}$ is no longer negligible on the operational timescale, and the martingale proof underpinning Born's rule in Sec.~\ref{sec:measurement} no longer applies. Operationally, the coarse-grained probability dynamics acquires an additional deterministic component: when colored fluctuations are treated as the physically natural Stratonovich limit of smooth fields, the reduction to an effective It\^o description generates a systematic Wong--Zakai drift \cite{Gardiner2009,Oksendal2013}.

For a two-level quantum system described by a single probability coordinate $p\in[0,1]$, the homogenized It\^o form can be written as
\begin{equation}
\label{eq:colored_ito_generic}
\mathrm dp = A(p)\,\mathrm dt + \sqrt{B(p)}\,\mathrm dW_t,
\end{equation}
where $B(p)=(1-p)^2D_{\mathrm{eff}}$ retains the same multiplicative structure inherited from the collapse law, while the drift $A(p)$ is generated by the finite-memory correction. In our case the correction has the characteristic symmetry-breaking profile
\begin{equation}
\label{eq:WZ_drift_shape}
A(p)\ \propto\ p(1-p)(1-2p),
\end{equation}
so the process is no longer a martingale: $\mathbb E[\mathrm dp\mid\mathcal F_t]\neq 0$. This drift does not merely renormalize a decay rate; it steers trajectories toward one absorbing boundary before the stopping time is reached. Consequently, the absorption probability $u(p_0)$, determined by the backward equation
\begin{equation}
\label{eq:backward_colored}
	A(p)u'(p)+\frac12 B(p)u''(p)=0,\quad u(0)=0,\ u(1)=1,
\end{equation}
need not satisfy $u(p_0)=p_0$. The deviation $u(p_0)-p_0$ is precisely the Born-line bias plotted in Fig.~\ref{fig:colored_noise}B: it is negligible in spectra dominated by short-memory components, but becomes appreciable when the long-memory fraction is increased and when the spectral alignment is chosen to maximize the drift accumulated before absorption.

\subsection{Why collapse predicts this, but GKSL cannot}
A natural conceptual question arises: if both our collapse dynamics and the GKSL master equation originate from the same relativistic path integral (Sec.~\ref{sec:decoherence}), why does colored noise generate observable Born-line deviations in our framework, while standard decoherence theory remains blind to them?

The answer is not microscopic but statistical. Collapse is a trajectory-level theory with a stopping-time rule: a definite outcome is declared only when the stochastic evolution reaches an absorbing boundary. The relevant observable is therefore a nonlinear first-passage quantity, namely the hitting probability of a boundary. Long-memory noise can bias this quantity through the drift $A(p)$ accumulated before absorption.

By contrast, the GKSL equation, and likewise its linear non-Markovian generalizations with memory kernels, evolves the unconditional ensemble state $\bar\rho(t)$ at each fixed time after averaging over the noise record. This averaging removes the trajectory-level conditioning and, crucially, replaces the stopping-time selection by a convex mixture of uninterrupted evolutions. As a result, any measurement probability extracted from $\bar\rho(t)$ using a fixed POVM element $M$,
\begin{equation}
\label{eq:POVM_linear}
	P(t)=\Tr\!\big[\bar\rho(t)M\big],
\end{equation}
remains linear in the ensemble state and therefore cannot encode the boundary-hitting statistics of a first-passage process.

In this sense, Born-line deviations are not a higher-order refinement hidden inside decoherence theory. Rather, they are a signature of record-conditioned dynamics with absorbing boundaries---a statistical structure that ensemble master equations are not designed to represent.

\subsection{Directed collapse and the resource-efficient quantum technology}
Once colored noise is treated as a physical feature of the environment rather than merely an unwanted disturbance, the mechanism discussed above can be used in a constructive way.  If the noise has a short correlation time, the collapse time $\tau_{\mathrm{clps}}$ can be adjusted by changing the noise spectrum without altering the Born probabilities of the final outcomes.  This provides a practical way to control how quickly a system relaxes into a definite state—for example, enabling either rapid reset or long-lived storage depending on the experimental requirement.  When longer memory components are present, the stochastic evolution acquires a small drift relative to the purely symmetric diffusion associated with Markovian noise, and then biases the collapse-probability that obeys Born's rule.  In this sense, colored noise provides a mechanism for guiding the collapse process toward a chosen state.

This method of rapid reset or decoherence suppression does not rely on additional qubits or complex feedback protocols; therefore, it offers significantly higher efficiency compared to traditional decoherence mitigation strategies. In Fig. \ref{fig:colored_noise}C, we compare our approach with current mainstream decoherence-suppression methods using a resource-effectiveness map. The vertical axis of this map evaluates the Decoherence Suppression Index ($C_* \in [0,1]$), defined as $C_* = \sum_{\alpha} \Omega_\alpha c_\alpha S_\alpha$. Here, $S_\alpha$ is the suppression factor and $c_\alpha$ is the fractional coverage for a given error channel $\alpha$. The horizontal axis computes the Normalized Resource Index ($\mathcal{R}_* \in (0,1]$), defined as $\mathcal{R}_* = \sum_j w_j \tilde{r}_j$, which aggregates the min-max normalized overheads for additional physical qubits ($r_q$), control sequence bandwidth ($r_c$), measurement latency ($r_m$), and cryogenic power loads ($r_p$). $w_j$ is the normalized weight factor of resource consumption, and in our work, $w_j=1/4$ is the equal weight factor.

On this plane, many established approaches achieve strong suppression only by paying a substantial resource cost: high $C_*$ typically comes with a large $\mathcal R_*$. In contrast, our directed-collapse way exhibits a advantage in the low-overhead, high-suppression regime. In the ``partial'' mode, relatively modest spectral shaping and anisotropic alignment already produces a competitive suppression score, while keeping the resource index low (the representative point in Fig.~\ref{fig:colored_noise}C lies around $C_*\approx 0.53$ at $\mathcal R_*\approx 0.20$). In the ``complete'' mode, enhancing the long-memory asymmetry drives the drift to dominate before absorption, pushing the dynamics toward near-deterministic capture into the chosen pointer state; correspondingly the map places this regime at high effectiveness with an even smaller overhead (around $C_*\sim 0.94$ at $\mathcal R_*\sim 0.05$).

These findings indicates that our theory holds significant potential for applications in quantum state manipulation.

\section*{Discussion and Conclusion}
This work provides a first-principles account of quantum measurement, decoherence, and effective non-Hermitian phenomena within a single relativistic path-integral framework, thereby addressing the central questions raised in the Introduction. We show that a mathematically consistent single-particle relativistic path integral bridging the relativistic action and the Dirac equation not only exists but also is well motivated as a spacetime description of relativistic quantum dynamics. It exposes an intrinsic nonlocal structure of the relativistic propagator and shows how this same structure can be activated by non-differentiable noise, thereby turning the axioms of quantum measurement from postulates into dynamical consequences. In the view of this theory, unitary evolution and quantum measurement are not separate rules but are described by the same evolution operator shown in Eq.~\ref{eq:shortprop_main}. The boundary between these regimes is depicted in Fig.~\ref{fig:unification}: a successful measurement requires an apparatus with sufficient temperature $T$, adequate electromagnetic noise strength (e.g., large $\sigma_E$ or $\sigma_B$ in the relevant band), or sufficient energy discrimination (large $E_{\mathrm{gap}}$). Consequently, collapse is not an instantaneous phenomenon but a rapid physical process. Its characteristic duration is set by the mean first-passage time of a bounded-martingale SDE, and is consistent at the order-of-magnitude level with representative experimental dephasing times.

Furthermore, our work directly addresses the physical origin of definite outcomes. In our theory, the driver is neither an external observer nor an added postulate, but the non-differentiable electromagnetic noise of the environment itself. Physically, such non-differentiability arises from the superposition of many individually smooth field modes sustained by thermal motion and charge-current fluctuations associated with ongoing quantum evolution and collapse. Collapse is therefore induced by environmental degrees of freedom, while each collapse event, in turn, contributes to the fluctuating background experienced by other subsystems. This establishes a closed, self-consistent loop: quantum-state collapse both depends on and feeds back into the noise bath that sustains it. In this sense, the state of our world is determined by the world itself, rather than by the choices of God.

Beyond resolving the measurement problem, this work places standard decoherence theory and effective non-Hermitian mechanics on a unified, first-principles foundation. When the environmental noise record is averaged, our stochastic dynamics reproduce the exact GKSL master equation, ensuring full compatibility with conventional open-system theory. However, the derivation clarifies that ensemble master equations represent a statistical coarse-graining that inherently traces out the underlying nonlinear trajectory dynamics. Similarly, the theory demonstrates that effective non-Hermitian Hamiltonians are not independent physical postulates. Instead, their algebraic structure arises as a conditioned, record-resolved representation of local relativistic noise. This distinction preserves the relativistic causal structure and resolves the superluminal signaling paradoxes often associated with non-Hermitian models, proving that such macroscopic boundary phenomena are statistical footprints of local noise conditioning rather than fundamental modifications of quantum kinematics.

Perhaps the most significant implication of our work is that it reframes environmental noise from a nuisance into a control resource. Our formula suggests that non-Markovian noise can induce directed collapse. Therefore, by engineering the spectral density and spatial anisotropy of the environmental noise, quantum states can be continuously steered and reset. Benchmarking this approach against established decoherence-suppression strategies (Fig.~\ref{fig:colored_noise}C) reveals that directed collapse can achieve high-fidelity state preparation with minimal hardware and cryogenic overhead, offering a low-cost alternative for scalable quantum control.  

An important next step is to examine multipartite entangled systems, in order to determine how directed collapse interacts with nonlocal correlations and to clarify the spatiotemporal dynamics of entanglement. In parallel with this theoretical extension, experimental verification is essential. To this end, we outline concrete and testable protocols in the Supplementary Materials, providing a clear pathway to validate the predicted collapse scalings across concrete physical platforms. In sum, our theory opens new ground in both principle and practice, providing a new perspective for understanding quantum theory and new methods for harnessing it.

\appendix
\section{From the Time-Slice Path Integral to the Evolution Equations}
\label{app:evolution_from_timeslice}

The main text derives the spinor and scalar evolution equations from the covariant time-slice path integral. A step-by-step derivation in fixed spatial dimensions ($n=1,2,3$), with the same integral identities checked case-by-case, was presented in our earlier paper~\cite{quantum7040059}. Here we give a dimension-uniform route that makes transparent (i) why the relativistic core integral reduces to a Bessel-type identity, and (ii) how the resulting closed form yields the evolution operator $e^{-\frac{\mathrm{i}}{\hbar}\hat H_f\epsilon}$, hence the relativistic wave equations.  We then summarize how smooth electromagnetic potentials enter through minimal coupling, and we close by proving the instantaneous identity kernel $K_L(\vec r_0+\Delta\vec r,t^+;\vec r_0,t)=\delta(\Delta\vec r)$ that is used in Appendix~\ref{app:covariant_propagator_perturb}.

We first consider the free scalar case ($V=0$, $\vec A=\vec 0$).  In the time-slice form of the main text,
\begin{equation}
	\begin{aligned}
		\Psi(\vec r,t+\epsilon)&=\int K_L(\vec r,t+\epsilon;\vec r_0,t)\,\Psi(\vec r_0,t)\,\mathrm d^n\vec r_0 \\
		&=\int\hat{L}_n^{-1} \underbrace{K (\vec{r},t+\epsilon;\vec{r}_0,t)\,L_n}_{K_G(\vec{r}-\vec{r}_0;\epsilon)}\Psi(\vec{r}_0,t)\mathrm{d}^n\vec{r}_0
	\end{aligned}
\label{eq:app_timeslice_map}
\end{equation}
the integral kernel depends only on the displacement $\Delta\vec r:=\vec r-\vec r_0$, hence we define
\begin{equation}
	K(\vec r,t+\epsilon;\vec r_0,t)L_n=K_G(\Delta\vec r,\epsilon),
\label{eq:app_translation_invariance}
\end{equation}
for notational convenience in the subsequent derivations. Then, we can see that \eqref{eq:app_timeslice_map} is a convolution on $\mathbb R^n$:
\begin{equation}
	\Psi(\vec r,t+\epsilon)=\hat{L}_n^{-1}[\bigl(K_G(\cdot,\epsilon)\ast \Psi(\cdot,t)\bigr)(\vec r)].
\end{equation}
Testing this map on a plane wave $\Psi_{\vec p}(\vec r_0,t)=e^{\frac{\mathrm{i}}{\hbar}\vec p\cdot \vec r_0}$ gives
\begin{equation}
	\begin{aligned}
		&\Psi_{\vec p}(\vec r,t+\epsilon)=\hat{L}_n^{-1}\int K_G(\vec r-\vec r_0,\epsilon)\,e^{\frac{\mathrm{i}}{\hbar}\vec p\cdot \vec r_0}\,\mathrm d^n\vec r_0 \\
		&=\hat{L}_n^{-1}e^{\frac{\mathrm{i}}{\hbar}\vec p\cdot \vec r}\int K_G(\Delta\vec r,\epsilon)\,e^{-\frac{\mathrm{i}}{\hbar}\vec p\cdot \Delta\vec r}\,\mathrm d^n\Delta\vec r.
	\end{aligned}
\label{eq:app_plane_wave_multiplier}
\end{equation}
Hence the action of the kernel on plane waves is fully discribed by the Fourier transform
\begin{equation}
	\widetilde K_G(\vec p,\epsilon):=\int K_G(\Delta\vec r,\epsilon)\,e^{-\frac{\mathrm{i}}{\hbar}\vec p\cdot \Delta\vec r}\,\mathrm d^n\Delta\vec r,
\label{eq:app_KL_fourier_def}
\end{equation}
so that \eqref{eq:app_plane_wave_multiplier} reads
\begin{equation}
	\Psi_{\vec p}(\vec r,t+\epsilon)=\hat{L}_n^{-1}\widetilde K_G(\vec p,\epsilon)\,e^{\frac{\mathrm{i}}{\hbar}\vec p\cdot \vec r}.
\label{eq:app_KL_action_def}
\end{equation}
Equation~\eqref{eq:app_KL_action_def} is the Fourier statement behind the time-slice evolution: once we show
\begin{equation}
	\widetilde K_G(\vec p,\epsilon)\!=\!\hat{L}_ne^{-\frac{\mathrm{i}}{\hbar}H_f(\vec p)\epsilon},
	\ \ \  H_f(\vec p)=\!\sqrt{m_0^2c^4+c^2\vec p^{\,2}},
\label{eq:app_goal_multiplier}
\end{equation}
we immediately obtain the scalar relativistic evolution $\mathrm i\hbar\,\partial_t\Psi=\sqrt{m_0^2c^4+c^2\hat{\vec p}^{\,2}}\,\Psi$ by the standard argument that the Fourier multiplier of the propagator is $e^{-\frac{\mathrm{i}}{\hbar}H_f\epsilon}$.

We now compute \eqref{eq:app_KL_fourier_def} from the explicit time-slice kernel induced by the relativistic action and the covariant transport prefactor of the main text.  Write $\vec v=(\vec r-\vec r_0)/\epsilon=\Delta\vec r/\epsilon$ and $\gamma_L(\vec v)=(1-\vec v^{\,2}/c^2)^{-1/2}$.  In the scalar case, the covariant transport reduces to the scalar factor
\begin{equation}
	L_n(\vec{v})=\gamma_L^{n/2}\left(\frac{\gamma_L+1}{2}\right)^{-\frac{n-2}{2}},
\label{eq:app_Lnr_identity}
\end{equation}
where the last equality is an algebraic identity (it is useful because the factor on the right is exactly the one appearing in the main-text integrand).

For the free scalar slice, the classical action is $S_r(\vec v)=-m_0c^2\sqrt{1-\vec v^{\,2}/c^2}\,\epsilon$ on timelike segments, together with the analytic continuation prescribed in the main text on the spacelike domain $|\vec v|>c$ (needed to keep the time-slice integral well defined).  Denote the resulting (continued) phase factor uniformly by $e^{\frac{\mathrm{i}}{\hbar}S_r(\vec v)}$.  Then $K_G$ has the structure
\begin{equation}
	K_G(\Delta\vec r,\epsilon)=C_0\,L_n(\vec v)\,e^{\frac{\mathrm{i}}{\hbar}S_r(\vec v)},\qquad \vec v=\frac{\Delta\vec r}{\epsilon},
\label{eq:app_KL_timeslice_structure}
\end{equation}
where $C_0 = (\frac{m_0}{2\mathrm i \pi \hbar (t-t_0)})^{n/2}$ is the normalization constant.

Substituting \eqref{eq:app_KL_timeslice_structure} into \eqref{eq:app_KL_fourier_def} and changing variables $\Delta\vec r=\epsilon\vec v$ yields
\begin{equation}
	\widetilde K_G(\vec p,\epsilon)=C_0\,\epsilon^n\int_{\mathbb R^n}L_n(\vec v)\,e^{\frac{\mathrm{i}}{\hbar}S_r(\vec v)}e^{-\frac{\mathrm{i}}{\hbar}\epsilon\,\vec p\cdot \vec v}\,\mathrm d^n\vec v.
\label{eq:app_multiplier_v}
\end{equation}
Because the integrand depends on $\vec v$ only through $|\vec v|$ except for the plane-wave factor, we pass to $n$-dimensional spherical coordinates $|\vec v|=\rho$:
\begin{equation*}
	\mathrm d^n\vec v=\rho^{n-1}\,\mathrm d\rho\,\mathrm d\Omega_{n-1},\qquad\vec p\cdot\vec v=|\vec p|\,\rho\,\cos\theta.
\end{equation*}
The angular integration is the standard $n$-dimensional Fourier--Bessel reduction:
\begin{equation}
	\int e^{-\mathrm{i}\kappa\rho\cos\theta}\,\mathrm d\Omega_{n-1}=(2\pi)^{n/2}(\kappa\rho)^{1-\frac n2}j_{\frac n2-1}(\kappa\rho),
\label{eq:app_spherical_bessel}
\end{equation}
where $\kappa:={\epsilon|\vec p|}/{\hbar}$ and $j_\alpha$ is the Bessel functions of the first kind. Applying \eqref{eq:app_spherical_bessel} to \eqref{eq:app_multiplier_v} gives a purely radial integral,
\begin{equation}
	\begin{aligned}
		\widetilde K_G(\vec p,\epsilon)&= C_0\epsilon^n(2\pi)^{n/2}\kappa^{1-\frac n2} \\
		&\int_0^\infty \rho^{\frac{n}{2}}\,L_n(\rho)\,e^{\frac{\mathrm{i}}{\hbar}S_r(\rho)}\,J_{\frac{n}{2}-1}(\kappa\rho)\,\mathrm d\rho.
	\end{aligned}
\label{eq:app_multiplier_radial}
\end{equation}
At this point one can proceed in two equivalent ways. By a hyperbolic change of variables (equivalent to $\rho/c=\sinh u$ on the continued contour and $x=\cosh u\in[1,\infty)$), the combination of the transport factor \eqref{eq:app_Lnr_identity} and the Jacobian reorganizes \eqref{eq:app_multiplier_radial} into an integral on $x\in[1,\infty)$ of the canonical form
\begin{equation}
	\int_1^\infty\left(\frac{x-1}{x+1}\right)^{\frac{\mu}{2}}e^{-\alpha x}J_{\mu}\Bigl(\beta\sqrt{x^2-1}\Bigr)\,\mathrm dx,
\label{eq:app_GR_form}
\end{equation}
with parameters
\begin{equation}
	\alpha=i\chi,\ \ \beta=iz,\ \ \chi:=\frac{m_0c^2\epsilon}{\hbar},\ \ z:=\frac{|\vec p|c\,\epsilon}{\hbar},
\label{eq:app_chi_z_def}
\end{equation}
and $\mu=\frac{n}{2}-1$. The closed form then follows directly from Gradshteyn \& Ryzhik Eq.~6.646.1~\cite{Gradshteyn2007}.
\begin{equation}
	\begin{aligned}
		&\int_1^\infty\left(\frac{x-1}{x+1}\right)^{\mu/2}e^{-\alpha x}J_{\mu}\!\bigl(\beta\sqrt{x^2-1}\bigr)\,\mathrm{d}x \\
		=&\frac{1}{\sqrt{\alpha^2+\beta^2}}\left(\frac{\beta}{\alpha+\sqrt{\alpha^2+\beta^2}}\right)^{\mu}e^{-\sqrt{\alpha^2+\beta^2}} .
	\end{aligned}
\end{equation}
Using \eqref{eq:app_chi_z_def}, one has
\begin{equation}
	\sqrt{\chi^2+z^2}=\frac{\epsilon}{\hbar}\sqrt{m_0^2c^4+c^2\vec p^{\,2}}=\frac{\epsilon}{\hbar}H_f(\vec p).
\label{eq:app_sqrt_relation}
\end{equation}
Substituting \eqref{eq:app_chi_z_def}--\eqref{eq:app_sqrt_relation} back into the multiplier formula \eqref{eq:app_multiplier_radial} yields
\begin{equation*}
	\begin{aligned}
		\widetilde K_G(\vec p,\epsilon)
			&=\frac{(m_0c^2+H_f(\vec p))^{\frac{2-n}{2}}}{2H_f(\vec p)}e^{-\frac{\mathrm{i}}{\hbar}H_f(\vec p)\epsilon} \\
			&=\hat{L}_ne^{-\frac{\mathrm{i}}{\hbar}H_f(\vec p)\epsilon}.
	\end{aligned}
\label{eq:app_multiplier_prefactor}
\end{equation*}
Inserting this into \eqref{eq:app_KL_action_def} recovers the scalar evolution equation
\begin{equation*}
	\mathrm i\hbar\,\partial_t\Psi(\vec r,t)=\sqrt{m_0^2c^4+c^2\hat{\vec p}^{\,2}}\,\Psi(\vec r,t),
\end{equation*}
as stated in the main text.

For spin-$\frac{1}{2}$ particles, the main text uses the Dirac-linear slice action and the spinor boost factor $F_B(\vec v)$.  The essential algebraic point is that the spinor phase is obtained by conjugating the scalar phase componentwise:
\begin{equation*}
	e^{\frac{\mathrm{i}}{\hbar}S_R(\vec v)}=F_B(\vec v)\,e^{\frac{\mathrm{i}}{\hbar}\beta S_r(\vec v)}\,F_B(\vec v)^{-1},
\end{equation*}
while the remaining scalar factors in the transport are the same dimension-matching prefactors already fixed in the scalar case.  Therefore the spinor time-slice integrals reduce to the scalar Bessel integral evaluated above, applied to each spinor block, and one obtains
\begin{equation*}
	\Psi(\vec r,t+\epsilon)=\exp\!\left[-\frac{\mathrm{i}}{\hbar}\Bigl(\beta m_0c^2+c\,\vec\alpha\cdot\hat{\vec p}\Bigr)\epsilon\right]\Psi(\vec r,t),
\end{equation*}
hence the Dirac equation $\mathrm i\hbar\,\partial_t\Psi=(\beta m_0c^2+c\vec\alpha\cdot\hat{\vec p})\Psi$.

Furthermore, when $V(\vec r,t)$ and $\vec A(\vec r,t)$ are differentiable, the time-slice can be evaluated to first order in $\epsilon$. The scalar-potential contribution gives the multiplicative phase $e^{-\frac{\mathrm{i}}{\hbar}qV(\vec r,t)\epsilon}$.  The vector potential enters through the standard line-integral phase in the kernel,
\begin{equation*}
	\exp\left(\frac{\mathrm{i} q}{\hbar}\int_{\vec r_0}^{\vec r}\vec A(\vec r_1,t)\cdot \mathrm d\vec r_1\right).
\end{equation*}
From the preceding analysis, we know that for any square-integrable function $\Psi'$, the following identity holds:
\begin{equation*}
	\begin{aligned}
		&\int K_G(\Delta\vec{r},\epsilon) \Psi'(\vec{r}_0) \mathrm{d}^n\Delta\vec{r} \\
		\equiv & L_n\bigg(\frac{\mathrm{i}\hbar\nabla_{\vec{r}_0}}{m_0}\bigg)\exp\left( \frac{\mathrm{i}\epsilon}{\hbar} \sqrt{m_0^2 c^4 - \hbar^2 c^2 \nabla^2_{\vec{r}_0}} \right) \Psi'(\vec{r}_0) \bigg|_{\vec{r}0=\vec{r}}.
	\end{aligned}
\end{equation*}
By setting $\Psi'(\vec{r}_0) = e^{\frac{\mathrm{i} q}{\hbar} \int_{\vec{r}_0}^{\vec{r}} \vec{A} \cdot \mathrm{d}\vec{r}_1} \Psi(\vec{r}_0)$, and invoking the operator identity:
\begin{equation}
\label{eq:app_A_smooth}
\begin{aligned}
	&-\mathrm{i}\hbar\nabla_{\vec{r}_0} \left[ e^{\frac{\mathrm{i} q}{\hbar} \int_{\vec{r}_0}^{\vec{r}} \vec{A} \cdot \mathrm{d}\vec{r}_1} \Psi(\vec{r}_0) \right]\bigg|_{\vec{r}_0=\vec{r}} \\
	=& \bigl( -\mathrm{i}\hbar\nabla_{\vec{r}} - q\vec{A}(\vec{r}, t) \bigr) \Psi(\vec{r}),
	\end{aligned}
\end{equation}
we can deduce that in the presence of a vector potential, one obtains the substitution $\hat{\vec{p}} \to \hat{\vec{p}} - q\vec{A}(\vec{r}, t)$ within $\hat{H}_f$. This leads to the Hamiltonian:
\begin{equation*}
	\hat{H} = \hat{H}_f(\hat{\vec{p}} - q\vec{A}) + qV,
\end{equation*}
which applies to both scalar and spinor cases, as stated in Eqs. (Dirac) and (scalar) of the main text.

It should be emphasized that Eq.~\eqref{eq:app_A_smooth} is not the gauge-conjugation identity $e^{\mathrm i\chi}\hat{\vec p}e^{-\mathrm i\chi}=\hat{\vec p}-q\vec A$, which would require $\nabla\chi=q\vec A$ and hence, in general, an integrability condition such as $\nabla\times\vec A=0$. Here $\chi$ is instead the open line integral along the time-slice segment, $\Theta_A(\vec r,\vec r_0)=\int_{\vec r_0}^{\vec r}\vec A\cdot d\vec r_1$, which may be path-dependent when $\nabla\times\vec A\neq0$. Equation~\eqref{eq:app_A_smooth} nevertheless remains valid because only the derivative with respect to the initial endpoint is needed, evaluated at the coincident-point limit $\vec r_0=\vec r$, for which $\nabla_{\vec r_0}\Theta_A|_{\vec r_0=\vec r}=-\vec A(\vec r,t)$. Thus Eq.~\eqref{eq:app_A_smooth} is a local endpoint identity rather than a special gauge-transformation construction.

Finally, we prove the distributional identity used in the perturbative-noise analysis. Assume that $\vec A(\vec r,t)$ and $V(\vec r,t)$ are differentiable at time $t$, so that the minimal-coupling reduction derived above is valid. Define the instantaneous covariant kernel by
\begin{equation*}
	\Psi(\vec r,t+\epsilon)=\int K_L(\vec r+\Delta\vec r,t+\epsilon;\vec r,t)\,
	\Psi(\vec r_0,t)\,\mathrm d^n\vec r_0 .
\end{equation*}
From the preceding derivation, its Fourier multiplier is
\begin{equation*}
	\hat L_n^{-1}\widetilde K_G(\vec p,\epsilon)=\exp\!\left[-\frac{\mathrm i}{\hbar}H(\vec p)\epsilon\right],
\end{equation*}
where
\begin{equation*}
	H(\vec p)=H_f(\vec p-q\vec A)+qV
\end{equation*}
is the smooth-field Hamiltonian obtained by minimal coupling. Therefore, in the instantaneous limit $\epsilon\to0^+$,
\begin{equation*}
	\hat L_n^{-1}\widetilde K_G(\vec p,\epsilon)\longrightarrow 1.
\end{equation*}
Hence the Fourier transform of the instantaneous kernel is identically equal to $1$. In the sense of tempered distributions, the unique kernel with Fourier transform $1$ is the Dirac delta. We thus obtain
\begin{equation}
	K_L(\vec r+\Delta\vec r,t^+;\vec r,t)=\delta(\Delta\vec r).
\label{eq:app_delta_claim}
\end{equation} 

\textit{Remark:}
The reduction in Eq.~\eqref{eq:app_A_smooth} requires temporal differentiability of $\vec A(t)$. If $\vec A$ is not differentiable---for instance, if it contains a jump in time---the argument leading to Eq.~\eqref{eq:app_delta_claim} no longer holds. On a single time slice we parametrize the straight segment by $\vec r_1(s)=\vec r_0+s(\vec r-\vec r_0)$ and $t_1(s)=t+s\epsilon$, $s\in[0,1]$.  
The phase entering the kernel reads
\begin{equation*}
	\Theta_A(\vec r,\vec r_0)=\int_{\vec r_0}^{\vec r}\vec A(t_1)\cdot\mathrm d\vec r_1=(\vec r-\vec r_0)\cdot\int_{0}^{1}\vec A(t+s\epsilon)\,\mathrm ds .
\end{equation*}
When $\vec A(t)$ is differentiable, the integral over $s$ may be expanded at $t$ and the derivative $-\mathrm i\hbar\nabla_{\vec r_0}$ acting on $e^{\frac{\mathrm i q}{\hbar}\Theta_A}\Psi(\vec r_0)$ reduces to the substitution $\hat{\vec p}\to \hat{\vec p}-q\vec A(\vec r,t)$. This is the mechanism behind Eq.~\eqref{eq:app_A_smooth} and the subsequent matching of the pullback factor with the time-slice convolution.

However, if $\vec A(t)$ is not differentiable, the average$\int_{0}^{1}\vec A(t+s\epsilon)\,\mathrm ds$ cannot be reduced to $\vec A(t)$ by a first-order expansion. For a jump in $\vec A(t)$, the integral samples both sides of the discontinuity, so that $\nabla_{\vec r_0}\Theta_A$ is not determined by a single endpoint value. Consequently, the operator identity underlying Eq.~\eqref{eq:app_A_smooth} fails in the ordinary sense: the action of $-\mathrm i\hbar\nabla_{\vec r_0}$ on the phase does not produce a unique $\vec A(\vec r,t)$.

Because the pullback operator at $t^+$ is evaluated at the endpoint field $\vec A(t^+)$, while the phase reduction involves a time-slice average of $\vec A$, the two structures no longer coincide. The identity $K_L(\vec r,t+\epsilon;\vec r_0,t)=\delta(\vec{r}-\vec{r}_0)$ therefore breaks down at non-smooth times. Non-differentiable gauge fields must be treated separately; the next appendix will discuss this situation.

\section{Perturbative Expansion of the Covariant Propagator}
\label{app:covariant_propagator_perturb}
In the previous appendix, we showed that for differentiable electromagnetic potentials the instantaneous covariant kernel reduces to the local identity distribution, Eq.~\eqref{eq:app_delta_claim}. When $\vec A(t)$ is no longer continuous across the time slice, however, this identity no longer holds. The question is then: what replaces the delta kernel in this non-smooth case?

At first sight, the answer may seem ambiguous. Indeed, if $\vec A(t)$ has a jump at the slicing time, the phase factor in the kernel can no longer be reduced by evaluating $\nabla_{\vec r_0}\Theta_A$ at a single endpoint value. In the actual time-slice calculation, however, the propagator is not evaluated pointwise. Rather, one Fourier expands the wavefunction into plane waves, applies the convolution kernel, and then performs the inverse Fourier transform. For a jump discontinuity, the inverse Fourier reconstruction selects the midpoint value of the discontinuity under the usual Dirichlet-Jordan conditions \cite{Zygmund2002,Katznelson2004,SteinShakarchi2003}. Therefore, the phase factor generated by the slice is governed by the midpoint field $\frac{\vec A(t^+)+\vec A(t^-)}{2}$, whereas the pullback operator in the covariant kernel is still evaluated at the endpoint field $\vec A(t^+)$. This mismatch is the entire source of the nontrivial correction.

Below we derive the resulting kernel for a spin-$\tfrac{1}{2}$ particle. To match the notation of the main text, we define the jump increment $\mathrm d\vec A_I:=\vec A(t^+)-\vec A(t^-)$, so that the midpoint field becomes
\begin{equation*}
	\frac{\vec A(t^+)+\vec A(t^-)}{2}=\vec A(t^+)-\frac12\,\mathrm d\vec A_I.
\end{equation*}
Since the following derivation is concerned mainly with the dependence on $\vec A$, it is convenient to introduce
\begin{equation*}
	\hat Q_n(\vec A):=\hat L_n=L_n\!\bigl(\vec\alpha\cdot\frac{\hat{\vec p}-q\vec A}{m_0}\bigr),
\end{equation*}
together with the bare time-slice kernel
\begin{equation*}
	\mathcal K_0(\Delta\vec r;\vec A):=\left(\frac{-m_0 c\beta}{2\pi\hbar|\Delta\vec r|}\right)^{n/2}e^{\frac{\mathrm i}{\hbar}\beta\,\vec\alpha\,m_0 c|\Delta\vec r|}e^{\frac{\mathrm i}{\hbar}q\vec A\cdot\Delta\vec r}.
\end{equation*}
In coordinate representation, $\hat Q_n^{-1}$ acts on the first spatial argument of the kernel.

For a differentiable background, the previous appendix established the identity
\begin{equation}
\label{eq:app_identity_delta_readable}
	\hat Q_n^{-1}(\vec A)\,\mathcal K_0(\Delta\vec r;\vec A)=\delta(\Delta\vec r).
\end{equation}
In the present non-smooth case, the phase factor is determined by the midpoint field $\vec A(t^+)-\tfrac12\,\mathrm d\vec A_I$, whereas the pullback operator still uses $\vec A(t^+)$. Therefore,
\begin{align}
	&K_L(\vec r+\Delta\vec r,t^+;\vec r,t)\nonumber\\
	=&\hat Q_n^{-1}\!\big(\vec A(t^+)\big)\mathcal K_0\left(\Delta\vec r;\vec A(t^+)-\tfrac{1}{2}\mathrm d\vec A_I\right)\nonumber\\
	=&\Big[\hat Q_n^{-1}\!\big(\vec A(t^+)\big)\hat Q_n\left(\vec A(t^+)-\tfrac12\,\mathrm d\vec A_I\right)\Big]\delta(\Delta\vec r)\nonumber\\
	=&\big[\mathbf 1+\delta\hat{\mathcal T}\big]\delta(\Delta\vec r),
\label{eq:app_KL_deltaT_readable}
\end{align}
where
\begin{equation}
\label{eq:app_deltaT_def_readable}
	\delta\hat{\mathcal T}:=\hat Q_n^{-1}\!\big(\vec A(t^+)\big)\,\hat Q_n\!\left(\vec A(t^+)-\tfrac12\,\mathrm d\vec A_I\right)-\mathbf 1.
\end{equation}

We now expand \eqref{eq:app_deltaT_def_readable} to first order in $\mathrm d\vec A_I$ in the same weak-momentum regime used in the main text, $\langle(\vec\alpha\cdot\hat{\vec P}_0)^2\rangle\ll m_0^2c^2$. Let $\hat{\vec{P}}=\hat{\vec{p}}-q\vec{A}$. As in the main text, we have
\begin{equation*}
	\hat Q_n(\vec{A})=F_B(\vec{\alpha}\cdot\tfrac{\hat{\vec{P}}}{m_0})\, F_n(\tfrac{\hat{\vec P}}{m_0})), \qquad  F_n(\hat{\vec{v}})={F}_V^{n}(\hat{\vec{v}}){F}_\Psi(\hat{\vec{v}}),
\end{equation*}
Consequently, the core of evaluating Eq.~\eqref{eq:app_deltaT_def_readable} involves the operation between two sets of operators ${F}_B^{-1}(\cdot){F}_B(\cdot)$ and ${F}_n^{-1}(\cdot){F}_n(\cdot)$. The boost contribution can be linearized because $\operatorname{arctanh}(x)=x+O(x^3)$.
With
\begin{equation*}
	\hat F_B=F_B(\beta\vec{\alpha}\cdot\tfrac{\hat{\vec P}}{m_0})=\exp\!\left[-\frac12\,\operatorname{arctanh}\!\left(\frac{\beta\,\vec\alpha\cdot\hat{\vec P}}{m_0c}\right)\right],
\end{equation*}
one finds
\begin{equation}
\label{eq:app_QB_mismatch_readable}
	\begin{aligned}
		&\hat{F}_B^{-1}(t^+)\hat{F}_B(t) \\
		=&F_B^{-1}(\vec{\alpha}\cdot\tfrac{\hat{\vec{P}}(t^+)}{m_0})\hat{F}_B(\vec{\alpha}\cdot\tfrac{\hat{\vec{P}}(t^+)-\frac{1}{2}q\vec{A}_I}{m_0}) \\
		=&\mathbf 1-\frac{q}{4m_0c}\,\beta\,\vec\alpha\cdot \mathrm d\vec A_I+O\!\left((\mathrm d\vec A_I)^2\right).
	\end{aligned}
\end{equation}
This term is anti-Hermitian at leading order and does not generate the population-separating diffusion channel emphasized in the main text; we keep it implicit from here on and focus on the Hermitian sector coming from the scalar mismatch.

The scalar factor depends on $\hat{\vec P}$ through the Lorentz-factor operator
\begin{equation}
	\hat\gamma_L=\left(1+\frac{(\vec\alpha\cdot\hat{\vec P})^2}{m_0^2c^2}\right)^{-1/2},
\end{equation}
and, in the notation of the main text,
\begin{equation}
	F_n(\tfrac{\hat{\vec{P}}}{m_0})=\left(\frac{1+\hat\gamma_L}{2}\right)^{-\frac{n-1}{2}}\hat\gamma_L^{\,n/2}.
\end{equation}
A logarithmic first-order expansion gives
\begin{equation}
\label{eq:app_F_ratio_readable}
	\begin{aligned}
		&F_n^{-1}(\tfrac{\hat{\vec P}(t^+)}{m_0})F_n(\tfrac{\hat{\vec P}(t^+)-\frac{1}{2}q\vec{A}_I}{m_0}) \\
		=&\mathbf 1+\left.\frac{\mathrm d}{\mathrm d\gamma_L}\ln F_n\right|_{\gamma_L=\hat\gamma_L}\,\mathrm d\hat\gamma_L+O\!\left((\mathrm d\hat\gamma_L)^2\right),
	\end{aligned}
\end{equation}
where
\begin{equation}
\label{eq:app_Lambda_def_readable}
	\left.\frac{\mathrm d}{\mathrm d\gamma}\ln F_n\right|_{\gamma_L=\hat{\gamma}_L}=-\frac{n-1}{2(1+\hat{\gamma}_L)}+\frac{n}{2\hat{\gamma}_L} \equiv \frac{1}{\Lambda_n(\hat H_0)}.
\end{equation}
Here $\Lambda_n(\hat H_0)$ is the same scalar operator function used in the main text.  Since the increment is assumed small compared with the background field, $\Lambda_n$ is evaluated on the background Hamiltonian $\hat H_0$ at time $t$, and corrections from the noise-dependence of $\Lambda_n$ are $O((\mathrm d\vec A_I)^2)$.

It remains to compute $\mathrm d\hat\gamma_L$ induced by $\hat{\vec P}=\hat{\vec P}_0-q\,\mathrm d\vec A_I$. Set $\hat X=(\vec\alpha\cdot\hat{\vec P})^2/(m_0^2c^2)$ so that $\hat\gamma_L=(1+\hat X)^{-1/2}$. Differentiation gives
\begin{equation*}
	\begin{aligned}
		&\mathrm d\hat\gamma_L=-\frac{1}{2}(1+\hat X)^{-\tfrac{3}{2}}\,\mathrm d\hat X,\\ 
		&\mathrm d\hat X=\frac{1}{m_0^2c^2}\bigl\{\vec\alpha\cdot\hat{\vec P},\ \vec\alpha\cdot\mathrm d\hat{\vec P}\bigr\}.
	\end{aligned}	
\end{equation*}
In the weak-momentum regime $(1+\hat X)^{-3/2}=\mathbf 1+O(\hat X)$ and $\mathrm d\hat{\vec P}=-q\,\mathrm d\vec A_I$, hence
\begin{equation}
\label{eq:app_dgamma_readable}
	\mathrm d\hat\gamma_L=\frac{q}{2m_0^2c^2}\bigl\{\vec\alpha\cdot\hat{\vec P}_0,\ \vec\alpha\cdot\mathrm d\vec A_I\bigr\}+O\!\left((\mathrm d\vec A_I)^2\right).
\end{equation}
Substituting \eqref{eq:app_dgamma_readable} into \eqref{eq:app_F_ratio_readable} and keeping the Hermitian sector yields the linear mismatch relevant for the main-text diffusion channel,
\begin{equation}
\label{eq:app_deltaT_H_readable}
	\delta\hat{\mathcal T}_{\mathrm H}=\frac{q\bigl\{\vec\alpha\cdot\hat{\vec P}_0,\ \vec\alpha\cdot\mathrm d\vec A_I\bigr\}}{8m_0^2c^2\,\Lambda_n(\hat H_0)}+O\!\left((\mathrm d\vec A_I)^2\right).
\end{equation}

To rewrite \eqref{eq:app_deltaT_H_readable} in a form that matches the main text, we use the standard Dirac identity \cite{BjorkenDrell1964}
\begin{equation*}
	(\vec\alpha\cdot\vec a)(\vec\alpha\cdot\vec b)=\vec a\cdot\vec b+\mathrm i\,\vec\Sigma\cdot(\vec a\times\vec b),
\end{equation*}
together with the commutator $[\hat p_j,f(\vec r)]=-\mathrm i\hbar\,\partial_j f(\vec r)$.  A direct expansion gives
\begin{align}
	&\bigl\{\vec\alpha\cdot\hat{\vec P}_0,\ \vec\alpha\cdot\mathrm d\vec A_I\bigr\} \nonumber \\
	=&\big(\hat{\vec P}_0\cdot\mathrm d\vec A_I+\mathrm d\vec A_I\cdot\hat{\vec P}_0\big)
	+\mathrm i\,\vec\Sigma\cdot\big(\hat{\vec P}_0\times\mathrm d\vec A_I+\mathrm d\vec A_I\times\hat{\vec P}_0\big)\nonumber\\
	=2&\,\hat{\vec P}_0\cdot\mathrm d\vec A_I-\mathrm i\hbar\,\nabla\cdot\mathrm d\vec A_I+\hbar\,\vec\Sigma\cdot(\nabla\times\mathrm d\vec A_I).\nonumber
\label{eq:app_anticom_expand_readable}
\end{align}
The divergence term is anti-Hermitian and therefore does not contribute to the Hermitian diffusion channel.
Writing $\mathrm d\vec B_I=\nabla\times\mathrm d\vec A_I$ and introducing the spin magnetic moment
\begin{equation*}
	\vec\mu_s=\frac{q\hbar}{2m_0}\vec\Sigma,
\end{equation*}
Eq.~\eqref{eq:app_deltaT_H_readable} becomes
\begin{equation*}
\label{eq:app_deltaT_final_readable}
	\delta\hat{\mathcal T}_{\mathrm H}=\left(\frac{q\,\mathrm d\vec A_I\cdot\hat{\vec P}_0}{4m_0^2c^2\,\Lambda_n(\hat H_0)}+\frac{\mathrm d\vec B_I\cdot\vec\mu_s}{4m_0c^2\,\Lambda_n(\hat H_0)}\right)+O\!\left((\mathrm d\vec A_I)^2\right),
\end{equation*}
and then yields the instantaneous covariant kernel to first order,
\begin{align}
	&K_L(\vec r_0+\Delta\vec r,t^{+};\vec r_0,t) \nonumber \\
	=&\delta(\Delta\vec r)+\left(\mathrm d\vec A_I\cdot\hat{\vec N}+\mathrm d\vec B_I\cdot\hat{\vec M}\right)\delta(\Delta\vec r),
\label{eq:app_KL_final_readable}
\end{align}
with
\begin{equation*}
	\hat{\vec N}:=\frac{q\,\hat{\vec P}_0}{4m_0^2c^2\,\Lambda_n(\hat H_0)},
\qquad
\hat{\vec M}:=\frac{\vec\mu_s}{4m_0c^2\,\Lambda_n(\hat H_0)}.
\end{equation*}
Equation~\eqref{eq:app_KL_final_readable} is Eq.~\eqref{eq:Nondiff_pathintegra}.  In a smooth field, $\mathrm d\vec A_I=O(\mathrm dt)$ and the correction disappears in the instantaneous limit.  In the non-smooth case used in the main text, the increment survives at the It\^o scaling order and becomes the leading stochastic contribution \cite{Gardiner2009,Oksendal2013}.

\section{Stochastic time-slice expansion of the covariant propagator}
\label{app:stochastic_timeslice_U}

This appendix explains how the infinitesimal propagator used in the main text, namely Eq.~\eqref{eq:shortprop_main}, arises from the covariant transport structure of the relativistic path integral together with the It\^o scaling of non-smooth field increments \cite{Gardiner2009,Oksendal2013}.

In the time-slice construction of the main text, the covariant propagation over one infinitesimal interval separates into (i) the usual unitary drift generated by the Hamiltonian and (ii) a transport mismatch coming from the pullback/pushforward factor evaluated at slightly different times. For ease of exposition, we adopt the notation from the previous chapter and use $\hat{Q}_n^{-1}$ to denote the pullback mapping. Then, the exact time-slice operator can be written in the form
\begin{equation}
\label{eq:app_U_factor}
	\hat{U}(t+\mathrm{d}t,t)=\underbrace{\hat{Q}_n^{-1}\!\big(\vec A\big)\,\hat{Q}_n\!\big(\vec A-\frac{1}{2}\mathrm{d}\vec{A}_I\big)}_{\hat{\mathcal M}(t+\mathrm{d}t,t)}\,\exp\!\left[-\frac{\mathrm{i}}{\hbar}\hat{H}(t)\,\mathrm{d}t\right],
\end{equation}
where $\hat{H}(t)=\hat{H}_f(\hat{\vec p}-q\vec A(t))+qV(t)$ is the Hamiltonian appearing in the main text (Dirac for spinors, square-root for scalars). If $\vec A(t)$ is differentiable, then $\mathrm{d}\vec{A}_I=\vec{A}(t^+)-\vec{A}(t)=0$ and the mismatch $\hat{\mathcal M}$ reduces to $\mathbf{1}+O(\mathrm{d}t)$; the time slice is then dominated by the unitary factor. When $\vec A(t)$ is not differentiable, the same mismatch factor is the only place where an $O(\sqrt{\mathrm{d}t})$ contribution can enter, and it is precisely this contribution that drives the stochastic dynamics in the main text.

To obtain the diffusive limit used in the collapse and decoherence sections, we idealize the field increment over $\mathrm{d}t$ by an It\^o scaling,
\begin{equation*}
	\mathrm{d}W_{\mu}=O(\sqrt{\mathrm{d}t}),\quad
	\mathbb E[\mathrm{d}W_{\mu}]=0,\quad
	\mathrm{d}W_{\mu}\mathrm{d}W_{\nu}=G_{\mu\nu}\,\mathrm{d}t,
\end{equation*}
with a constant positive semidefinite covariance matrix $G=[G_{\mu\nu}]\succeq 0$ \cite{Gardiner2009,Oksendal2013}.
In this regime the mismatch operator $\hat{\mathcal M}$ admits an expansion
\begin{equation}
\label{eq:app_M_expand}
	\hat{\mathcal M}(t+\mathrm{d}t,t)=\mathbf{1}+\delta\hat{\mathcal M}^{(1)}+\delta\hat{\mathcal M}^{(2)}+o(\mathrm{d}t).
\end{equation}
The first-order term is linear in the increment and can therefore be written as a sum over independent Wiener channels,
\begin{equation}
\label{eq:app_M1}
	\delta\hat{\mathcal M}^{(1)}=\sum_{\mu}\hat{J}_{\mu}\,\mathrm{d}W_{\mu}.
\end{equation}
The operators $\hat{J}_{\mu}$ are the (Hermitian) couplings derived from the first-order variation of $\hat{Q}_n^{-1}\!\big(\vec A\big)\,\hat{Q}_n\!\big(\vec A-\frac{1}{2}\mathrm{d}\vec{A}_I\big)$ in the perturbative analysis of the covariant propagator (Appendix~\ref{app:covariant_propagator_perturb}) and are the same couplings that enter the stochastic master equations in the main text \cite{WisemanMilburn2009,JacobsSteck2006}.

Once \eqref{eq:app_M1} is fixed, the $O(\mathrm{d}t)$ term in \eqref{eq:app_M_expand} is not free: products of the $O(\sqrt{\mathrm{d}t})$ increments contribute at order $\mathrm{d}t$ through the quadratic variation $\mathrm{d}W_{\mu}\mathrm{d}W_{\nu}$. Keeping only terms that survive at order $\mathrm{d}t$, the It\^o bookkeeping gives
\begin{equation}
\label{eq:app_M2}
	\delta\hat{\mathcal M}^{(2)}=-\frac{1}{2}\sum_{\mu,\nu}\hat{J}_{\nu}\hat{J}_{\mu}\,\mathrm{d}W_{\mu}\mathrm{d}W_{\nu}=-\frac{1}{2}\sum_{\mu,\nu}G_{\mu\nu}\,\hat{J}_{\nu}\hat{J}_{\mu}\,\mathrm{d}t.
\end{equation}
This is the familiar It\^o correction term: it is the same mechanism that produces the Lindblad dissipator after ensemble averaging, and it is required for the averaged short-time map to be completely positive and trace-preserving \cite{Breuer2002,WisemanMilburn2009}.

The unitary drift is expanded in the standard way,
\begin{equation}
\label{eq:app_unitary_expand}
	\exp\!\left[-\frac{\mathrm{i}}{\hbar}\hat{H}(t)\,\mathrm{d}t\right]=\mathbf{1}-\frac{\mathrm{i}}{\hbar}\hat{H}(t)\,\mathrm{d}t+o(\mathrm{d}t).
\end{equation}
Substituting \eqref{eq:app_M_expand}--\eqref{eq:app_unitary_expand} into \eqref{eq:app_U_factor} and multiplying out, the cross term
$\big(\sum_{\mu}\hat{L}_{\mu}\mathrm{d}W_{\mu}\big)\big(-\tfrac{\mathrm{i}}{\hbar}\hat{H}\mathrm{d}t\big)$
is of order $\mathrm{d}t^{3/2}$ and is discarded in the $\mathrm{d}t\to 0$ time-slice limit.
Collecting all contributions up to order $\mathrm{d}t$ yields
\begin{align}
	\hat{U}(t+\mathrm{d}t,t)=&\mathbf{1}-\frac{\mathrm{i}}{\hbar}\hat{H}(t)\,\mathrm{d}t+\sum_{\mu}\hat{J}_{\mu}\,\mathrm{d}W_{\mu} \nonumber\\
	&-\frac{1}{2}\sum_{\mu,\nu}G_{\mu\nu}\,\hat{J}_{\nu}\hat{J}_{\mu}\,\mathrm{d}t
+o(\mathrm{d}t),\nonumber
\end{align}
which is exactly Eq.~\eqref{eq:shortprop_main}.

\section{Estimation of the characteristic frequency $Z_{\rm cr}$ (zero-crossing rate scale)}
\label{app:zcr}

In the main text we introduced a characteristic frequency $Z_{\rm cr}$ in order to parametrize the
magnetic-noise increment as $\mathrm d\vec B_I \sim Z_{\rm cr}\,\sigma_B\,\mathrm d\vec W_t$.
The purpose of this appendix is to (i) state a mathematically standard definition of a
zero-crossing (level-crossing) rate for a stationary random field, and (ii) provide a physically
transparent estimate for $Z_{\rm cr}$ that is consistent with the scaling used in the main text.

Consider first a single Cartesian component $B(t)$ of a stationary, zero-mean random magnetic field. Assume that $B(t)$ is a stationary Gaussian process that is mean-square differentiable, so that $\langle \dot B(t)^2\rangle < \infty$. Let
\begin{equation*}
	R(\tau):=\langle B(t)B(t+\tau)\rangle=\int_{-\infty}^{\infty} e^{\mathrm i\omega\tau}\,S_B(\omega)\,\frac{\mathrm d\omega}{2\pi},
\end{equation*}
where $S_B(\omega)$ is the (two-sided) power spectral density. Rice's classical level-crossing theory gives the expected zero-crossing rate (the mean number of crossings of $B(t)$ through $0$ per unit time) as
\begin{equation*}
	\begin{aligned}
		&\nu_0=\frac{1}{\pi}\sqrt{\frac{-R''(0)}{R(0)}}=\frac{1}{\pi}\sqrt{\frac{w_2}{w_0}} \\
		& w_k:=\int_{-\infty}^{\infty}\omega^k S_B(\omega)\,\frac{\mathrm d\omega}{2\pi}.
	\end{aligned}
\label{eq:app_rice}
\end{equation*}
Equivalently, since $m_0=\langle B^2\rangle$ and $m_2=\langle \dot B^2\rangle$, one may write
\begin{equation*}
	\nu_0=\frac{1}{\pi}\frac{\sqrt{\langle \dot B^2\rangle}}{\sqrt{\langle B^2\rangle}}.
\end{equation*}
We therefore define the zero-crossing frequency scale
\begin{equation}
	Z_{\rm cr}:=\pi\nu_0=\sqrt{\frac{w_2}{w_0}}=\frac{\sqrt{\langle \dot B^2\rangle}}{\sqrt{\langle B^2\rangle}},
\label{eq:app_zcr_def}
\end{equation}
which quantifies the typical inverse time scale of field fluctuations. The vector case can be handled componentwise (or by replacing $B^2$ with $\tfrac13|\vec B|^2$ under isotropy), and the role of $Z_{\rm cr}$ in the main text is precisely to provide such an effective inverse time scale.

If one idealizes the environment as infinitely broadband (a ``white-noise'' limit), then $S_B(\omega)$ is flat as $|\omega|\to\infty$, and the second spectral moment $m_2=\int \omega^2 S_B(\omega)\,\mathrm d\omega/(2\pi)$ diverges. In that case $\nu_0$ and $Z_{\rm cr}$ are not finite, reflecting the well-known fact that ideal white noise is not mean-square differentiable. Equation~\eqref{eq:app_zcr_def} therefore makes explicit why the magnetic channel must be coarse-grained by a finite bandwidth (or a finite correlation time) before it can be used as an effective Wiener-driven increment in the It\^o description.

A simple illustration is a band-limited spectrum $S_B(\omega)=S_0\,\mathbf 1_{|\omega|<\omega_c}$. Then $m_0=2S_0\omega_c/(2\pi)$ and $m_2=2S_0\omega_c^3/(3\cdot 2\pi)$, hence
\begin{equation*}
	Z_{\rm cr}=\sqrt{\frac{m_2}{m_0}}=\frac{\omega_c}{\sqrt{3}},\qquad\nu_0=\frac{\omega_c}{\pi\sqrt{3}}.
\label{eq:app_bandlimited}
\end{equation*}
Thus $Z_{\rm cr}$ is, up to an order-unity factor, the effective high-frequency cutoff of the noise.

To obtain a closed-form estimate suitable for order-of-magnitude modelling, we parametrize the effective cutoff by a carrier frequency scale and a dimensionless modulation factor. We take the carrier scale to be the Compton frequency (written with $h$), $f_C:=\frac{m_0c^2}{h}$, and write
\begin{equation}
	Z_{\rm cr}=f_C\,Z_1,
\label{eq:app_zcr_factor}
\end{equation}
where $Z_1$ summarizes the enhancement of sign changes induced by thermal collisions and local charge-current fluctuations in the environment.

A minimal dimensionally consistent estimate is to take $Z_1$ proportional to the thermal velocity ratio $v_{\rm th}/c\sim \sqrt{k_BT/(m_0c^2)}$, multiplied by the square of an effective number $N_0$ of independent charged degrees of freedom contributing to the local field fluctuations:
\begin{equation}
	Z_1\sim C_1\,N_0^2\,\frac{v_{\rm th}}{c}
	\sim C_1\,N_0^2\sqrt{\frac{k_BT}{m_0c^2}},
\label{eq:app_z1_model}
\end{equation}
with $C_1=O(1)$ absorbing non-universal geometric and material factors.

To estimate $N_0$ we use a cross-sectional (not closed-surface) flux argument at the relativistic length scale $L_*:=h/(m_0c)$. We consider a field magnitude $|\vec E|$ such that the work scale across $L_*$ is comparable to the relativistic momentum-energy scale,
\begin{equation*}
	q|\vec E|L_*\sim m_0c^2\quad\Rightarrow\quad|\vec E|\sim \frac{m_0c^2}{qL_*}=\frac{m_0^2c^3}{q h}.
\label{eq:app_Escale}
\end{equation*}
The associated cross-sectional electric flux through an area $\sim \pi L_*^2$ is then
\begin{equation*}
	\Phi_E \sim \pi L_*^2|\vec E|\sim \pi\left(\frac{h}{m_0c}\right)^2\frac{m_0^2c^3}{q h}=\frac{\pi h c}{q}.
\label{eq:app_flux}
\end{equation*}
Interpreting $\epsilon_0\Phi_E$ as an effective charge scale driving the local field fluctuations,
the corresponding effective number of elementary charges is
\begin{equation}
	N_0\sim \frac{\epsilon_0\Phi_E}{q}=\frac{\pi h c\epsilon_0}{q^2}=\frac{\pi}{2\alpha_{\rm EM}},
\label{eq:app_N0}
\end{equation}
where $\alpha_{\rm EM}=q^2/(4\pi\epsilon_0\hbar c)$ is the fine-structure constant.

Combining Eqs.~\eqref{eq:app_zcr_factor}--\eqref{eq:app_N0} yields the estimate used in the main text:
\begin{equation}
	Z_{\rm cr}\sim C_1\,\frac{\pi^2 h c^2\epsilon_0^2}{q^4}\sqrt{k_BT\,m_0c^2}.
\label{eq:app_zcr_final}
\end{equation}
In the numerical estimates in the main text we set $C_1=1$. The role of
Eq.~\eqref{eq:app_zcr_final} is to supply an effective inverse correlation time for the magnetic
channel when an idealized It\^o description is adopted; it should be interpreted as an
order-of-magnitude estimate whose prefactor may be renormalized by the actual spectral density,
geometry, and material properties of the experimental environment.

\begin{acknowledgements}
We thank Dr. Stefan Nimmrichter (Universit\"at Siegen), Professors Pingxing Chen and Wei Wu (National University of Defense Technology), Professor Ming Gong (University of Science and Technology of China) for their helpful discussions on theoretical aspects. We are also grateful to Yi Xie and Jie Zhang (National University of Defense Technology) for discussions on experimental design. 

The authors declare no competing financial interests. This work is supported by the National Natural Science Foundation of China (Grant No. 11904099), the Natural Science Foundation of Hunan Province of China (Grant No. 2021JJ30210), and the Excellent Y outh Program of Hunan Provincial Department of Education (Grant No.22B0609).
\end{acknowledgements}

\end{document}